\documentclass[twocolumn]{aastex631} 
\usepackage{threeparttable}
\usepackage{multirow}
\usepackage{times}
\usepackage{enumitem}
\usepackage{natbib}
\usepackage{url} 
\usepackage{amsmath,bm}
\usepackage{multirow}
\usepackage[dvipsnames]{xcolor}
\usepackage[table]{xcolor}
\usepackage{tikz}
\usepackage{rotating}
\usepackage{ulem}
\usepackage{textcomp}
\definecolor{lightblue}{RGB}{220,230,241}  
\usepackage{fontawesome5}

\shorttitle{SpecCLIP: Aligning and Translating Spectroscopic Measurements for Stars} 
\shortauthors{Xiaosheng Zhao et al.}

\begin{document}

\title{SpecCLIP: Aligning and Translating Spectroscopic Measurements for Stars} 

\correspondingauthor{Yang Huang}
\email{huangyang@ucas.ac.cn}

\author[0000-0002-8328-1447]{Xiaosheng Zhao}
\altaffiliation{These authors contributed equally to this work.}
\affiliation{School of Astronomy and Space Science, University of Chinese Academy of Sciences, Beijing 100049, People's Republic of China}
\affiliation{National Astronomical Observatories, Chinese Academy of Sciences, Beijing 100012, People's Republic of China}
\affiliation{Department of Physics \& Astronomy, The Johns Hopkins University, Baltimore, MD 21218, USA}

\author[0000-0003-3250-2876]{Yang Huang}
\affiliation{School of Astronomy and Space Science, University of Chinese Academy of Sciences, Beijing 100049, People's Republic of China}
\affiliation{National Astronomical Observatories, Chinese Academy of Sciences, Beijing 100012, People's Republic of China}

\author[0009-0003-7834-1112]{Guirong Xue}
\altaffiliation{These authors contributed equally to this work.}
\affiliation{Zhejiang Laboratory, Hangzhou 311121, People's Republic of China}

\author[0000-0001-8011-8401]{Xiao Kong}
\altaffiliation{These authors contributed equally to this work.}
\affiliation{National Astronomical Observatories, Chinese Academy of Sciences, Beijing 100012, People's Republic of China}
\affiliation{School of Astronomy and Space Science, University of Chinese Academy of Sciences, Beijing 100049, People's Republic of China}

\author[0000-0002-2874-2706]{Jifeng Liu}
\affiliation{National Astronomical Observatories, Chinese Academy of Sciences, Beijing 100012, People's Republic of China}
\affiliation{School of Astronomy and Space Science, University of Chinese Academy of Sciences, Beijing 100049, People's Republic of China}

\author{Xiaoyu Tang}
\affiliation{Research Center for Astronomical Computing, Zhejiang Laboratory, Hangzhou 311121, People's Republic of China}

\author[0000-0003-4573-6233]{Timothy C. Beers}
\affiliation{Department of Physics and Astronomy, University of Notre Dame, Notre Dame, IN 46556, USA}
\affiliation{Joint Institute for Nuclear Astrophysics -- Center for the Evolution of the Elements (JINA-CEE), USA}

\author[0000-0001-5082-9536]{Yuan-Sen Ting}
\affiliation{Department of Astronomy, The Ohio State University, 140 West 18th Avenue, Columbus, OH 43210, USA}
\affiliation{Center for Cosmology and AstroParticle Physics (CCAPP), The Ohio State University, Columbus, OH 43210, USA}

\author[0000-0001-7865-2648]{A-Li Luo}
\affiliation{National Astronomical Observatories, Chinese Academy of Sciences, Beijing 100012, People's Republic of China}
\affiliation{School of Astronomy and Space Science, University of Chinese Academy of Sciences, Beijing 100049, People's Republic of China}

\begin{abstract}

In recent years, large language models (LLMs) have transformed natural language understanding through vast datasets and large-scale parameterization. Inspired by this success, we present SpecCLIP, a foundation model framework that extends LLM-inspired methodologies to stellar spectral analysis. Stellar spectra, akin to structured language, encode rich physical and chemical information about stars. By training foundation models on large-scale spectral datasets, our goal is to learn robust and informative embeddings that support diverse downstream applications. As a proof of concept, SpecCLIP involves pre-training on two spectral types--LAMOST low-resolution and Gaia XP--followed by contrastive alignment using the CLIP (Contrastive Language-Image Pre-training) framework, adapted to associate spectra from different instruments. This alignment is complemented by auxiliary decoders that preserve spectrum-specific information and enable \textit{translation} (prediction) between spectral types, the former being achieved by maximizing mutual information between embeddings and input spectra. The result is a cross-spectrum framework that enables intrinsic calibration and flexible applications across instruments. We demonstrate that fine-tuning these models on moderate-sized labeled datasets improves adaptability to tasks such as stellar-parameter estimation and chemical-abundance determination. SpecCLIP also enhances the accuracy and precision of parameter estimates bench-marked against external survey data. In addition, its similarity search and cross-spectrum prediction capabilities offer potential for anomaly detection. Our results suggest that contrastively trained foundation models enriched with spectrum-aware decoders can advance precision stellar spectroscopy. Our code {\tt SpecCLIP} is publicly available on \href{https://github.com/Xiaosheng-Zhao/SpecCLIP}{GitHub} {\faGithub}.

\end{abstract}

\section{Introduction}

Over the past decades, large-scale spectroscopic surveys have revolutionized our understanding of the formation and evolution of the Milky Way \citep{1989ARA&A..27..555G, 2002ARA&A..40..487F,10.1117/12.461524,2009IAUS..258...11W, 2020ARA&A..58..205H}. These advances have been driven by three key forces. First, the continuous development of large-scale spectroscopic surveys -- such as RAVE \citep{2006AJ....132.1645S}, SEGUE \citep{2010ApJ...714..663D}, APOGEE \citep{2017AJ....154...94M}, GALAH \citep{2015MNRAS.449.2604D}, LAMOST \citep{2012RAA....12..723Z}, and DESI \citep{2016arXiv161100036D} -- has provided an unprecedented volume of stellar spectra across diverse Galactic populations. Second, the creation of powerful data infrastructures \citep{1991ASSL..171...89H, doi:10.1126/science.293.5537.2037, 10.1117/12.461524, 10.1117/12.2057445, 2017A&A...605A..52M}, exemplified by the SkyServer \citep{2001cs.......11015S} and CasJobs \citep{2005cs........2072O} systems built for the Sloan Digital Sky Survey (SDSS, \citealp{1999RSPTA.357...93M, 2003AJ....126.2081A}, has democratized access to these datasets and enabled efficient large-scale analyses. Third, the refinement of algorithms for extracting physical parameters from these spectra has enabled increasingly precise stellar characterization.

The latter effort encompasses traditional line-index methods, such as the SEGUE Stellar Parameter Pipeline (SSPP) \citep{2008AJ....136.2022L}; template-matching techniques, including UlySS \citep{2009A&A...501.1269K}, the LAMOST stellar parameter pipeline \citep[LASP;][]{2014IAUS..306..340W} and the LAMOST stellar parameter pipeline at Peking University \citep[LSP3;][]{2015MNRAS.448..822X}; as well as a range of machine learning approaches, among which the SSPP also incorporates a neural network module, alongside methods such as the Cannon \citep{2015ApJ...808...16N}, the Payne \citep{2017ApJ...849L...9T, 2019ApJ...879...69T},the  DD-Payne \citep{, 2019ApJS..245...34X}, and the TransformerPayne \citep{2025ApJ...980...66R}. All of these approaches, though diverse in methodology, rely heavily on supervision -- either empirical or theoretical.

Empirical approaches are limited by the coverage of the reference libraries they use, even when these libraries are derived from fundamental measurements. For instance, the official LAMOST stellar parameter pipeline \citep[LASP;][]{2014IAUS..306..340W}, which is based on the UlySS algorithm, can only measure iron abundances down to [Fe/H]~$=-2.5$, due to the limited parameter coverage of the ELODIE library \citep{2004PASP..116..693M}. Theoretical approaches, while offering broader parameter coverage, are still subject to discrepancies between synthetic and observed spectra.

In addition, most existing pipelines are designed primarily to estimate atmospheric parameters, such as effective temperature ($T_{\rm eff}$), surface gravity (log~$g$) and iron abundance ([Fe/H]), in conjunction with a small set of elemental abundances. In contrast, determining other stellar properties, including reddening, stellar mass, and age, typically requires dedicated, task-specific pipelines. These efforts are further complicated by the diversity of spectral data in terms of wavelength coverage, resolution, and signal-to-noise ratio (SNR) -- we refer to such quantities here as ``\textbf{modalities}". Achieving consistency and placing all inferred parameters onto a uniform scale remains a significant challenge in the prevailing framework, where each parameter is commonly derived using a distinct, often non-overlapping model.

In parallel, the past five years have witnessed the remarkable success of large language models (LLMs) in natural language understanding, conversational AI, and text generation \citep{NIPS2017_3f5ee243, 2018arXiv181004805D, radford2018improving, radford_language_2019, NEURIPS2020_1457c0d6}. Breakthroughs in high-impact scientific domains 
-- such as AlphaFold for protein structure prediction \citep{jumper2021highly} -- have been enabled by the combination of massive datasets, large-scale models, and modern computational infrastructure \citep{journals/corr/abs-2001-08361}.

Stellar spectra can be analogized to a structured language: their rich absorption features and overall shapes encode key information about a star’s physical properties and evolutionary history. With the accumulation of millions of stellar spectra, it has become feasible to train foundation models \citep{2024MNRAS.527.1494L,2024arXiv241016081B,2024MNRAS.531.4990P,2024arXiv241108842R,2024arXiv240514930S,2024arXiv241221130Z,2025arXiv250315312E,2025arXiv250101070P} on these data using techniques inspired by LLMs. Here, ``foundation models’’ refer to models pre-trained on large and diverse datasets. The broader the spectral distribution and parameter ranges used during pre-training, the better the model can learn the underlying structure of the ``spectral language''. Once pre-trained, these models can be quickly fine-tuned with a small set of high-quality labels to perform a variety of downstream tasks, such as parameter estimation.

A particularly promising framework for learning across modalities  is the Contrastive Language–Image Pre-training (CLIP) algorithm \citep{2021arXiv210300020R}, which aligns text and image representations via contrastive learning. CLIP jointly trains two encoders (one for each modality) by maximizing the similarity between representations of matched pairs and minimizing it for mismatched ones. The result is a shared embedding space that enables direct comparison between different modalities, allowing the model to retrieve or match one modality given the other, even for new inputs not seen during training.

When applied to stellar spectra \citep{2024arXiv241016081B,2024MNRAS.531.4990P,2024arXiv241108842R}, CLIP-style models can align spectra from different instruments or modalities with other astrophysical measurements, allowing for more flexible downstream tasks such as parameter estimation and anomaly detection. However, a known limitation of CLIP is that it prioritizes the shared information between modalities, potentially removing modality-specific features that are still important for some downstream tasks \citep{2024Entrp..26..252S}.

Partly motivated by this issue, we introduce the \textbf{SpecCLIP} project -- a unified framework for cross-modal representation learning of stellar spectra. SpecCLIP begins with pre-training on two types of spectra: LAMOST \citep{2012RAA....12.1197C} low-resolution spectra \citep[LRS;][]{2012RAA....12..723Z} and Gaia XP spectra \citep{2023A&A...674A...2D,2023A&A...674A...1G}. These are aligned in a shared embedding space using a CLIP-like contrastive objective, enhanced by auxiliary decoders to preserve the mutual information between the learned embeddings and the input spectra. 

Mutual information (MI, \citealp{10.5555/2981345.2981371, 2019arXiv190506922P, 2018arXiv180806670D,2023mla..confE..30S, 2025arXiv250612230T})\footnote{Formally, the mutual information between two continuous random variables \( X \) and \( Y \) is defined as
\[
I(X; Y) = \int\!\!\!\int p(x, y) \log \frac{p(x, y)}{p(x)p(y)}\,dx\,dy,
\]
where \( p(x, y) \) is the joint density, and \( p(x) \), \( p(y) \) are the marginals. For discrete variables, the integral becomes a sum.} is a key concept in information theory that quantifies how much one variable tells us about another. It serves as a natural objective for representation learning. Although we do not explicitly compute MI between the spectra and their embeddings, we adopt a simple and effective strategy \citep{2022arXiv220307004W}: increasing contrastive training with input reconstruction as a regularization mechanism, similar to using a decoder in an autoencoder to encourage informative representations.

Another interesting feature enabled by our model is spectrum-to-spectrum \textit{translation}. If the embeddings capture physically meaningful and shared information between spectra of different modalities, then it should be possible to predict one modality from the other, given suitable (spectra-to-spectra) supervision. In parallel, \citet{2024arXiv241016081B} demonstrated a similar idea using contrastive learning between Gaia XP and RVS spectra, employing CNN and multilayer perceptron (MLP) architectures. In our framework, we combine contrastive training and cross-modal prediction, augmented by spectral reconstruction within a unified architecture, enabling both spectrum-to-spectrum and spectrum-to-parameter applications.

The key contributions of this work are:
\begin{itemize}
    \item We design distinct tokenization and model structures tailored to LAMOST LRS and Gaia XP spectra to improve representation learning.
    \item We demonstrate that our unified model enables both in-modal and cross-modal search, as well as cross-modal prediction.
\end{itemize}

The remainder of this paper is organized as follows. In Section~\ref{sec:clip_intro}, we introduce the SpecCLIP model, including the separate pre-trained models and the CLIP-based alignment. Section~\ref{sec:downstream_tasks} describes the downstream tasks, including modeling and sample selection for parameter estimation. The results, including parameter inference, spectral retrieval, and cross-modal prediction, are presented in Section~\ref{sec:results}. Section~\ref{sec:discussion} offers further discussion and Section~\ref{sec:summary} summarizes the work. Additional material is provided in the appendices: Appendix~\ref{sec:pred_more} (additional results); Appendix~\ref{sec: normalization} (continuum fitting); Appendix~\ref{sec:nf} (normalizing flows for parameter inference); Appendix~\ref{sec:pre-trained} (pre-training details); Appendix~\ref{sec:proj_decode} (projection models and decoders); and Appendix~\ref{sec:hyper_summary} (summary of key hyper-parameters and configurations)

\begin{figure*}[hbt!]
\begin{center}
\includegraphics[scale=0.41,angle=0]{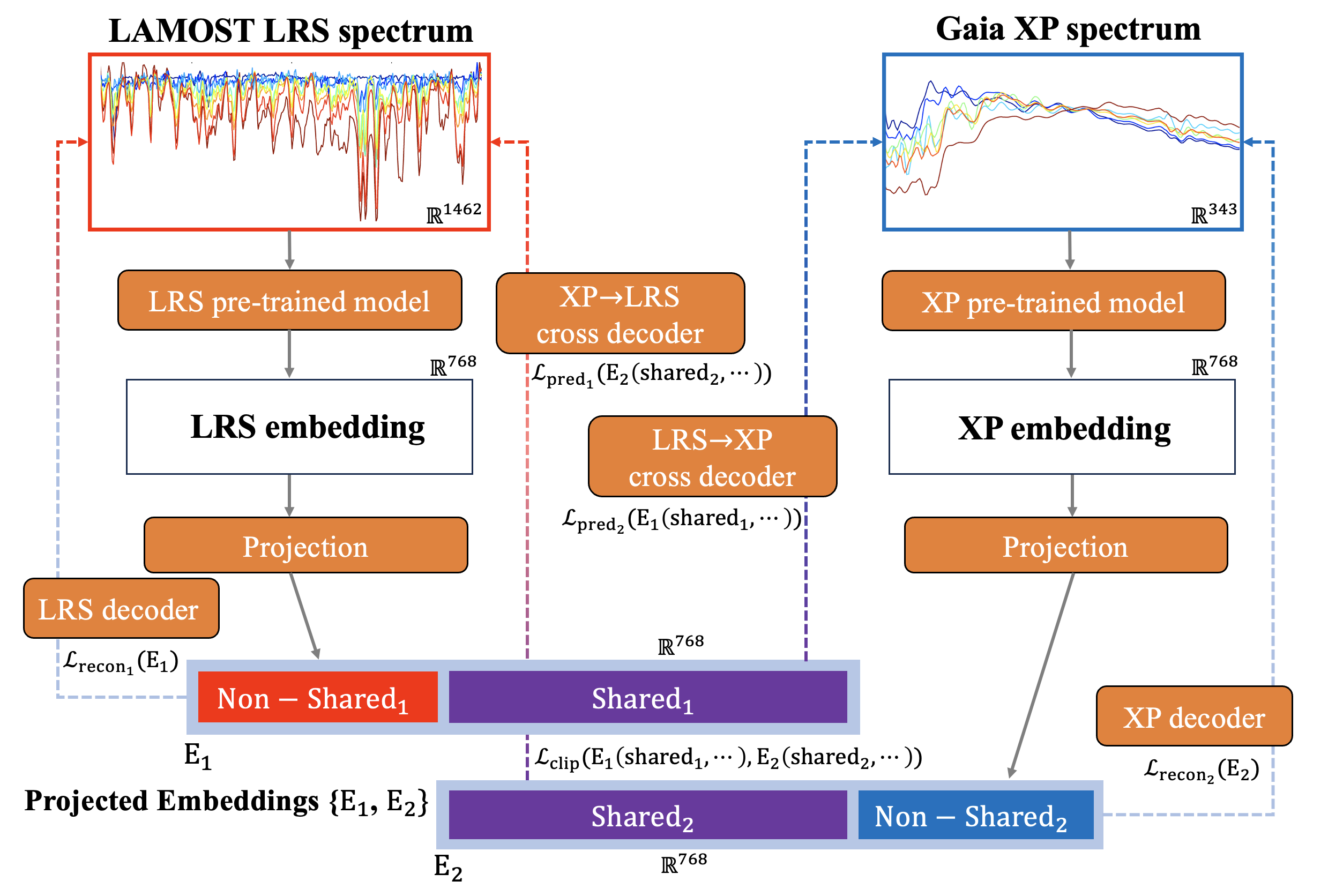} 
\caption{
Architecture of SpecCLIP. Two types of spectra (Shown here are examples of normalized LAMOST LRS and Gaia XP spectra varying with metallicities) are passed through two pre-trained spectral foundation models to obtain embeddings, where the pre-trained models can be either transformer-based networks or multilayer perceptron (MLP)-based autoencoders. These embeddings are then projected into a joint embedding space, which may optionally be split into a shared and a non-shared subspace. Based on the projected embeddings, we construct various loss functions to enable CLIP-like contrastive learning, cross-modal prediction, and spectral reconstruction. The combination of these loss functions results in five model variants: a baseline CLIP without decoders, CLIP-r with only reconstruction decoders, CLIP-p with only prediction decoders, CLIP-pr with full decoders, and CLIP-split with full decoders and an explicit separation of shared and non-shared embedding spaces (see Section~\ref{sec:variants})}
\end{center}
\label{fig:SpecCLIP_model}
\end{figure*} 

\section{SpecCLIP}
\label{sec:clip_intro}

SpecCLIP, developed in this work, is a foundation model designed to align stellar spectra across different modalities, such as those with varying wavelength coverage, resolution, and SNRs, using a CLIP-inspired architecture.\footnote{CLIP stands for Contrastive Language–Image Pre-training. In this work, we adopt the contrastive learning concept from the original CLIP framework, applying it to spectra–spectra pre-training. Although our use case differs from the original text–image alignment, we retain the term CLIP for consistency.} Our approach builds upon transformer-based foundation models or MLP-based autoencoders tailored for specific spectral types. These are combined with contrastive learning to align different spectra using a standard contrastive loss, supplemented by additional modules to capture both shared and modality-specific information. Once trained, SpecCLIP enables various downstream tasks with either branch (not combination) of modalities, using a relatively small number of labeled examples (i.e., few-shot learning, as referred to in the literature).

Figure~\ref{fig:SpecCLIP_model} provides an overview of the SpecCLIP framework. Foundation models are first independently pre-trained on each spectral modality in an unsupervised fashion. These models are then aligned using contrastive learning, with additional decoders added to enhance information retention. After this process, the final model is capable of handling a wide range of downstream tasks. Below, we describe the key experimental settings, including data-selection criteria, model architectures, loss functions for the foundation model training, and the connection of reconstruction loss to mutual information. Further details of the implementation are provided in Appendix~\ref{sec:pre-trained} and Appendix~\ref{sec:proj_decode}.

\subsection{Pre-trained Foundation Model for LAMOST LRS}

The most recent LAMOST data release (DR11)\footnote{\url{https://www.lamost.org/dr11/}} includes more than 10 million low-resolution spectra \citep{2015RAA....15.1095L}, used for a variety of scientific tasks, including estimation of stellar parameters, chemical abundances, reddening, radial velocity, stellar mass, and age. Our aim in pre-training a foundation model for LAMOST LRS is to learn informative and transferable representations that support many of these tasks with a small to moderate amount of labeled data.

In our pre-trained LRS model, each spectrum is tokenized into overlapping segments (tokens) of 20 flux points 20 flux points ($\approx 22$ \AA\ at a wavelength of $5000$\,\text{\AA}, corresponding to $\sim$10 resolution elements and capturing typical local line structures), with the overlap of 10 points to preserve continuity, producing 146 tokens per spectrum. These tokens, together with one special token--the logarithmic standard deviation of the spectrum--are passed through 6 self-attention layers, each yielding a 768-dimensional token embedding. During training, 6 non-overlapping chunks (each 10 tokens, corresponding to an overall effective masking rate of $\approx$\,45\%) are randomly masked to encourage robust representation learning. The resulting transformer-like framework with mask modeling (denoted as the masked transformer, or MT) has a total of 42.7 million trainable parameters.

\subsection{Pre-trained Foundation Model for Gaia XP Spectra}

Gaia XP low-resolution spectra \citep{2023A&A...674A...2D}, obtained via the Blue (with resolving power $R$ ranging from 30 to 100) and Red ($R$ ranging from 70 to 100) Photometers (BP and RP), are essential for determining key stellar properties such as stellar atmospheric parameters and chemical compositions. The differences in resolution and wavelength range compared to LAMOST LRS motivate the construction of a separate foundation model tailored to the XP modality.

Two types of models are explored for Gaia XP. The first is a transformer-based MT model, structurally similar to that used for LAMOST LRS, but with tokenization at the individual flux-point level, yielding 343 tokens per spectrum, together with two special tokens—the mean and standard deviation of the spectrum. Like the LRS model, this version uses 6 self-attention layers and masks 6 chunks (20 tokens each, corresponding to an overall masking rate of $\approx$\,35\%) during training.

The second model is a MLP-based ordinary autoencoder (denoted as OAE), with a bottleneck layer of 768 dimensions. Both XP models have approximately the same number of trainable parameters as the LRS model and are trained for the same number of epochs. In this paper, we use the OAE for Gaia XP to test results in tables, unless otherwise noted, because of its better performance, as discussed in Section~\ref{sec:transformer_vs_oae}.

\subsection{Contrastive Learning with Decoders}
\label{sec:variants}

To align the two spectral modalities, we use 820,568 paired LAMOST LRS and Gaia XP spectra. The backbone of the alignment model leverages contrastive loss, augmented by auxiliary decoders that contribute to additional supervision.

As shown in Figure~\ref{fig:SpecCLIP_model}, the embeddings of the LRS and XP foundation models are projected onto a shared embedding space. The LRS projection head includes a cross-attention block with a recurrence vector that is learnable, following \citet{2024MNRAS.531.4990P}. For Gaia XP, the projection head is either a cross-attention block (when following the attention-based model) or an MLP (when based on the OAE model). The core alignment objective is the contrastive loss between these projected embeddings.

To enrich the learned embeddings, we incorporate four auxiliary decoders:
\begin{itemize}
    \item Two in-modal decoders reconstruct each spectrum from its projected embedding;
    \item Two cross-modal decoders predict one spectrum modality from the other.
\end{itemize}
These components serve to retain modality-specific information, support cross-modal translation, and increase the robustness of learned representations. Although our multi-decoder design is motivated by the need to enrich spectral embeddings with diverse reconstruction pathways, we note that related ideas have appeared independently in other domains, such as the sensor-agnostic image retrieval framework in remote sensing proposed by \citet{2024arXiv240107782H}. As discussed in further detail in the final subsection of this section, reconstructing the original spectra also helps to increase the mutual information between the projected embeddings and the inputs.

\paragraph{Model Variants}
To assess the contributions of each model component, we construct five SpecCLIP variants:

\begin{itemize}
    \item \textbf{CLIP:} A baseline contrastive model without auxiliary decoders.
    \item \textbf{CLIP-r:} Adds only reconstruction decoders to the baseline.
    \item \textbf{CLIP-p:} Adds only cross-modal prediction decoders to the baseline.
    \item \textbf{CLIP-pr:} Adds both reconstruction and prediction decoders, implicitly encouraging shared and modality-specific representation learning.
    \item \textbf{CLIP-split:} Extends CLIP-pr by explicitly partitioning the embedding space into shared and modality-specific subspaces through two separate projection networks, potentially disentangling the two spaces.
\end{itemize}

\subsection{Loss Functions during Contrastive Training}
\label{sec:loss}

The total training objective $\mathcal{L}_{\text{total}}$ for the \texttt{SpecCLIP} model -- whether for CLIP or CLIP-variants -- comprises three components: a contrastive CLIP loss, a reconstruction loss, and a cross-modal prediction loss. The total loss is a weighted sum:

\begin{equation}
\mathcal{L}_{\text{total}} = \mathcal{L}_{\text{clip}} + \delta_{\text{recon}} \cdot w_{\text{recon}} \cdot \mathcal{L}_{\text{recon}} + \delta_{\text{pred}} \cdot w_{\text{pred}} \cdot \mathcal{L}_{\text{pred}},
\label{eq:loss}
\end{equation}

\noindent where $\delta_{\text{recon}}$ and $\delta_{\text{pred}}$ are binary indicators that control the inclusion of reconstruction losses and cross-modal prediction, respectively. The weights $w_{\text{recon}}$ and $w_{\text{pred}}$ control the relative contribution of the reconstruction and prediction losses and are fixed at 1 (i.e., equal weighting) throughout this work.

\paragraph{Contrastive CLIP Loss}

The contrastive loss aligns the XP and LRS embeddings in a shared embedding space. Let $f_{\text{xp}}$ and $f_{\text{lrs}}$ denote the encoders for XP and LRS inputs (including their pre-trained models and projection networks). For a batch of $N$ paired examples $\{(x_i^{\text{xp}}, x_i^{\text{lrs}})\}_{i=1}^N$, their projected embeddings are computed as described below.

In the CLIP-pr model, we use all projected embeddings and normalize them to unit length: $z_i^{\text{xp}} = f_{\text{xp}}(x_i^{\text{xp}})$ and $z_i^{\text{lrs}} = f_{\text{lrs}}(x_i^{\text{lrs}})$, where each embedding is L2-normalized. In the CLIP-split model, only the shared components of the embeddings are used and similarly normalized: $z_i^{\text{xp}} = f_{\text{xp}}^{\text{shared}}(x_i^{\text{xp}})$ and $z_i^{\text{lrs}} = f_{\text{lrs}}^{\text{shared}}(x_i^{\text{lrs}})$.

The similarity matrix are scaled by a temperature parameter $\tau$ (set to 15.5, following \citealt{2024MNRAS.531.4990P}):

\begin{align}
\ell_{i,j} = \tau \cdot \langle z_i^{\text{xp}}, z_j^{\text{lrs}} \rangle.
\end{align}

\noindent where $\langle \cdot, \cdot \rangle$ denotes the dot product between two embedding vectors. This dot product becomes equivalent to cosine similarity if the vectors are normalized. The parameter $\tau$ controls how confidently the contrastive loss distinguishes positives from negatives. The larger $\tau$ leads to greater confidence and greater push/pull between the matched and mismatched pairs.

Then the CLIP loss is computed as:

\begin{equation}
\mathcal{L}_{\text{clip}} = \frac{1}{2N} \sum_{i=1}^N \left[ \text{CE}(\ell_{i,:}, i) + \text{CE}(\ell_{:,i}^T, i) \right],
\end{equation}

\noindent where $\ell_{i,:}$ and $\ell_{:,i}$ are the similarity scores for XP-to-LRS and LRS-to-XP matching, respectively, and $\text{CE}(\cdot, i)$ denotes the cross-entropy loss with label $i$ defined as $\text{CE}(\ell_{i,:}, i) = -\log \left( \exp(\ell_{i,i}) / \sum_{j=1}^N \exp(\ell_{i,j}) \right)$. The cross-entropy loss encourages the matched pairs to have a higher similarity than the mismatched ones in the shared embedding space.

\paragraph{Reconstruction Loss}

Each modality has its own decoder to reconstruct the original spectrum from its embedding. In the CLIP-pr model, reconstruction is based on the full projected embedding. In the CLIP-r and CLIP-split models, reconstruction uses both shared and non-shared embeddings, which come from separate projection branches. Let $\hat{x}_i^{\text{xp}}$ and $\hat{x}_i^{\text{lrs}}$ denote the reconstructions. The loss is:

\begin{equation}
\mathcal{L}_{\text{recon}} = \frac{1}{N} \sum_{i=1}^N \left[ \|x_i^{\text{xp}} - \hat{x}_i^{\text{xp}}\|_1 + \left\| \frac{x_i^{\text{lrs}} - \mu_i}{\sigma_i} - \hat{x}_i^{\text{lrs}} \right\|_1 \right],
\end{equation}

\noindent where $\mu_i$ and $\sigma_i$ are the mean and standard deviation of the LRS input $x_i^{\text{lrs}}$, used to further normalize the input before reconstruction\footnote{This normalization step is optional. We initially adopted this design for potential training stability. Although each spectrum is already flux-normalized by continuum division during pre-processing, this additional normalization (1) centers the input distribution and (2) focuses the decoder on learning the shape of spectral features. This makes the projected embedding retain essential LRS information. In contrast, for Gaia XP, we reconstruct the relative fluxes (colors) normalized by the 550 nm flux, aiming to recover the full chromatic information.}.

\paragraph{Cross-Modal Prediction Loss}

To facilitate spectrum translation, cross-modal decoders predict one modality from the embedding of the other. In CLIP-pr, this is done by using the full embedding. In CLIP-p and CLIP-split, only the shared embedding component is used for cross-modal prediction. Let $\hat{x}_i^{\text{xp} \leftarrow \text{lrs}}$ and $\hat{x}_i^{\text{lrs} \leftarrow \text{xp}}$ be the cross-predicted spectra:

\begin{equation}
\mathcal{L}_{\text{pred}} = \frac{1}{N} \sum_{i=1}^N \left[ \|x_i^{\text{xp}} - \hat{x}_i^{\text{xp} \leftarrow \text{lrs}}\|_1 + \|x_i^{\text{lrs}} - \hat{x}_i^{\text{lrs} \leftarrow \text{xp}}\|_1 \right].
\end{equation}

\subsection{Connection of Reconstruction Loss to Mutual Information}
\label{sec:mi}

Minimizing the reconstruction loss encourages the latent embedding to retain as much information as possible about the original input spectra. This intuition can be formalized via the mutual information between the input spectrum (either $x^{\text{xp}}$ or $x^{\text{lrs}}$) and its corresponding embedding ($z^{\text{xp}}$ or $z^{\text{lrs}}$). For notational simplicity, we use $x$ and $z$ to denote a generic input spectrum and its embedding, respectively. The mutual information can be written as \citep{10.5555/2981345.2981371,2025arXiv250612230T}:
\begin{equation}
I(z, x) = H(x) - H(x|z),
\end{equation}
where $H(x) = -\mathbb{E}_{p(x)}[\log p(x)]$ is the marginal entropy of the spectrum and $H(x|z) = -\mathbb{E}_{p(z,x)}[\log p(x|z)]$ is the conditional entropy given the embedding. Since $H(x)$ is independent of model parameters, maximizing $I(z, x)$ is equivalent to minimizing $H(x|z)$.

However, the true conditional distribution $p(x|z)$ is typically intractable. In practice, we introduce a variational approximation $q(x|z)$—often realized by a neural decoder—leading to the following bound:
\begin{equation}
H(x|z) = -\mathbb{E}_{p(z,x)}[\log p(x|z)] \leq -\mathbb{E}_{p(z,x)}[\log q(x|z)],
\end{equation}
where the inequality follows from the non-negativity of the Kullback–Leibler divergence $\mathrm{KL}(p(x|z)\|q(x|z))$ \citep{10.5555/2981345.2981371,2019arXiv190506922P}. Thus, minimizing the expected negative log-likelihood under $q(x|z)$ serves as a variational lower bound on the mutual information $I(z, x)$.

When $q(x|z)$ is modeled as a Laplace distribution centered at a deterministic decoder output $\hat{x}(z)$, the negative log-likelihood reduces (up to constants) to the $\ell_1$ reconstruction error:
\begin{equation}
- \log q(x|z) \propto \|x - \hat{x}(z)\|_1 + c,
\end{equation}
which justifies the use of L1 loss in our reconstruction objective. This choice is particularly appropriate for stellar spectra, especially LAMOST LRS data, which contain sharp absorption features and may occasionally include unflagged bad pixels. The L1 loss is robust to such outliers and better preserves narrow spectral lines compared to L2.

Alternatively, assuming a Gaussian likelihood $q(x|z) = \mathcal{N}(x; \hat{x}(z), \sigma^2 I)$ leads to the standard mean squared error (MSE) loss:
\begin{equation}
- \log q(x|z) \propto \|x - \hat{x}(z)\|_2^2 + c,
\end{equation}
which penalizes larger residuals more strongly and encourages smooth reconstruction. While MSE may be suitable for high-SNR or denoised spectra, it tends to overly smooth localized features and is less robust to localized artifacts.

In summary, minimizing the reconstruction loss—whether L1 or L2—amounts to maximizing a variational lower bound on the mutual information between the input spectrum and its embedding. This encourages the representation to be informative and faithful to the original input.

\section{Downstream Tasks}
\label{sec:downstream_tasks}

We evaluate the performance of SpecCLIP on several downstream tasks, including parameter estimation, spectral retrieval (or search), and cross-modal prediction given a query spectrum. This section focuses on parameter estimation, detailing the model choices and sample-selection criteria.

\subsection{Models for Parameter Estimation}
\label{sec:downstream}

We explore two complementary approaches for stellar parameter estimation: MLPs and simulation-based inference (SBI), also known as implicit-likelihood or likelihood-free inference \citep{tejero-cantero2020sbi,2024OJAp....7E..54H}. SBI enables inference of parameter posteriors by learning the underlying distribution (e.g., the posterior itself) and evaluating on observed data, without requiring an explicit likelihood function.

\subsubsection{Multilayer Perceptrons}
\label{sec:mlp}

For most downstream tasks involving LAMOST LRS and Gaia XP spectra, we employ MLPs due to their training efficiency and scalability, which make them suitable for processing large datasets with limited computational resources. Each MLP has the following layer architecture: $[\text{input\_dim}, 1024, 512, 64, 1]$, where $\text{input\_dim} = \{1462, 343, 768\}$ depending on whether the input is raw LRS spectra, raw XP spectra, or embeddings. The output dimension is one, corresponding to a single stellar parameter. For pre-trained models via MT, we first reduce the representations by averaging over the sequence dimension (e.g., from $[146, 768]$ to $[768]$ for the LAMOST model). The resulting 768-dimensional embeddings and their high-quality labels are then used as input to the MLP. For CLIP models and the pre-trained XP model via OAE, no reduction of dimension is required and we directly use the 768-dimensional (projected) embeddings together with their labels. Each MLP model has approximately 1.4 million trainable parameters when applied to the embeddings.

\subsubsection{Simulation-Based Inference}
\label{sec:sbi}

For selected tasks, we also apply SBI for posterior inference. We report the median of the posterior as the point estimate.

We follow the Neural Posterior Estimation (NPE) framework to directly estimate the posterior distribution of parameters given observations, using neural density estimators. This NPE-based approach is relatively straightforward to implement and computationally efficient, making it a practical choice compared to other variants such as Neural Likelihood Estimation (NLE). We adopt two types of (conditional) normalizing flows as density estimators from the \texttt{sbi} package \citep{tejero-cantero2020sbi}: Masked Autoregressive Flow (MAF) \citep{2017arXiv170507057P} and Neural Spline Flow (NSF) \citep{2019arXiv190604032D}. Each SBI model uses two transformations with 60 hidden units (unless otherwise noted), totaling roughly 0.1 million trainable parameters. Details of the adopted normalizing flows are described in Appendix~\ref{sec:nf}.

To evaluate the calibration of the inferred posteriors, we perform simulation-based calibration (SBC, \citealp{talts2018validating}) using rank statistics, also known as Probability Integral Transform (PIT) values in some literature \citep{PITvalue}, complemented by the Kolmogorov–Smirnov (K–S) test \citep{Kolmogorov1992}. The K–S test measures the maximum discrepancy between the empirical cumulative distribution function (CDF) of the ranks and the expected uniform CDF, providing a non-parametric test of distributional consistency. For each (out of 200 in total) spectrum–parameter pair, we draw posterior samples and compute the rank of the ground-truth parameter value. Well-calibrated posteriors yield uniformly distributed ranks, which we assess via the K–S test. We consider results valid if the $p$-value exceeds 0.05, indicating that there is no significant deviation from uniformity and therefore reliable posterior coverage.

\section{Datasets}
\label{sec:datasets}
\subsection{Sample Selection for Pre-training}
\label{sec:sample_pretrain}
For the LRS model, we select a subset of 966,082 high-quality spectra for pre-training, using a 9:1 train–validation split (the same split ratio is used for the other modeling efforts described in the following two subsections). The selection criteria are: (1) SNR in the SDSS $g$-band greater than 50, and (2) apparent $g$-band magnitude less than 15.8. Additionally, we include all spectral types beyond the AFGK classes, while attempting to balance the four AFGK types themselves. We also aim to balance giant and dwarf stars; however, due to the intrinsic distribution of the dataset, the final ratio of dwarfs to giants is approximately 3.8:1. 

For the XP model, we pre-train this model using one million Gaia XP spectra, with around 80\% having matching LAMOST LRS spectra.
Each XP spectrum consists of 343 flux points spanning wavelengths from 336\,nm to 1021\,nm. 

\subsection{Sample Selection for Parameter Estimation}
\label{sec:downstream_selection}

For LAMOST LRS spectra, we evaluate parameter estimation across several classes of physical properties:

\begin{enumerate}
    \item Stellar atmospheric parameters: effective temperature ($T_{\rm eff}$), surface gravity (log~$g$), and iron abundance ([Fe/H]);
    \item Other elemental abundances: [$\alpha$/Fe], [C/Fe], [N/Fe], [Mg/Fe], [O/Fe], [Al/Fe], [Si/Fe], [Ca/Fe], [Ti/Fe], [Mn/Fe], [Ni/Fe], [Cr/Fe];
    \item Asteroseismic parameters and derived physical parameters: large frequency separation ($\Delta\nu$), frequency of maximum oscillation power ($\nu_{\max}$), stellar mass ($M_\odot$), radius ($R_\odot$), age (Gyr), and period spacing of gravity modes ($\Delta\Pi$);
    \item Other parameters: radial velocity ($v_r$) and extinction $E(BP-RP)$.
\end{enumerate}

We selected approximately 100,000 stars per parameter to balance the parameter distribution and computational tractability. The quality-control criteria include $g$-band spectral SNR$_g$ $>$ 20, Gaia $g$-band magnitude $<$ 16.5, and $|v_r| < 800$ km\,s$^{-1}$ to exclude likely extragalactic sources. LAMOST DR11 is used throughout; we adopt a $3''$ matching radius between it and other catalogs. For APOGEE, we typically (unless otherwise noted) require APOGEE SNR (median SNR per pixel in combined frame (at \texttt{apStar} sampling))$>$ 50.

Sample selection strategies for individual parameters are summarized below:

\begin{itemize}
    \item \textbf{Radial Velocity ($v_r$):} We select 100,299 stars in common between APOGEE DR17 \citep{2022ApJS..259...35A} and the LAMOST LRS sample. Their radial velocities are approximately uniformly distributed over the range $[-800, 800]$ km s$^{-1}$, divided into 800 bins, each containing up to 1600 stars.
    
    \item \textbf{Effective Temperature ($T_{\rm eff}$):} Approximately 100,000 stars are uniformly sampled in the $T_{\rm eff}$–$\log g$ plane, divided into a $100 \times 100$ grid, with up to four stars per bin. The values of $T_{\rm eff}$ are adopted from the LAMOST catalog.
    
    \item \textbf{Surface Gravity ($\log g$):} 100,000 stars combining log~$g$ labels from Kepler \citep[][mainly for red giant stars]{2022ApJ...927..167L} and APOGEE DR17 \citep{2022ApJS..259...35A}. Binned into 750 intervals (up to 350 stars per bin), with priority given to Kepler-based values.

    \item \textbf{Iron Abundance ($\mathrm{[Fe/H]}$):} A total of 100,118 stars are selected from a merged dataset comprising APOGEE DR17 \citep[][for stars with ${\rm [Fe/H]}>-2.0$]{2022ApJS..259...35A}, supplemented with metal-poor stars from the PASTEL and SAGA compilations \citep{2024ApJ...974..192H}, the LAMOST/Subaru VMP sample \citep{2022ApJ...931..147L}, and UMP datasets \citep{2019MNRAS.484.2166S}. A targeted sampling strategy is used to ensure balanced coverage in [Fe/H].

    \item \textbf{Other Elemental Abundances}: 85,400 stars from LAMOST–APOGEE cross-matches, filtered by APOGEE SNR $>$ 40 and valid abundance flags (i.e., \texttt{C\_FE\_FLAG} = 0, \texttt{N\_FE\_FLAG} = 0, \texttt{MG\_FE\_FLAG} = 0). A 2D binning in [Mg/Fe]–[Fe/H] space ($632 \times 632$ bins) ensures broad and uniform sampling. The [\(\alpha\)/Fe] values used here are computed from APOGEE measurements as \texttt{ALPHA\_M} $-$ (\texttt{FE\_H} $-$ \texttt{M\_H}), where \texttt{ALPHA\_M} includes O, Mg, Si, S, Ca, Ti, and \ion{Ti}{2} \citep{2020AJ....160..120J}.
    
    \item \textbf{Extinction ($E(BP-RP)$):} A total of 86,000 stars, with extinction values estimated using the star-pair technique \citep{2013MNRAS.430.2188Y} based on the LAMOST stellar parameter catalog.
    
    \item \textbf{Asteroseismic Parameters:} A total of 3,029 stars have asteroseismic parameters $\Delta \nu$ and $\nu_{\rm max}$ derived from Kepler light curves \citep{2022ApJ...927..167L, 2014ApJS..210....1C}, including 2,718 red giant stars and 311 main-sequence/turn-off stars. In addition, 4,034 red giant stars have measured gravity-mode period spacing $\Delta \Pi$, taken from published catalogs \citep{2016A&A...588A..87V}.
\end{itemize}

For Gaia XP spectra, we examine the following key parameters (with sample number in brackets) -- [$\alpha$/Fe] (94,584), [C/Fe] (99,934), [N/Fe] (95,089), $T_{\rm eff}$ (100,000), log~$g$, [Fe/H] (113,218), and color excess $E(BP-RP)$ (99,087) -- based on consistent data sources. These datasets are cross-matched with the Gaia XP catalog. To enhance [C/Fe] coverage in the metal-poor regime, we further include stars from the LAMOST very metal-poor catalog, where [C/Fe] has been estimated using a customized version of the SEGUE Stellar Parameter Pipeline \citep[LSSPP;][]{2015AJ....150..187L}.

All datasets are divided into a ratio of 0.81:0.09:0.10 for training, validation, and testing, respectively. The validation set is used for early stopping, that is, training is halted when performance on the validation set no longer improves within 10 training epochs. The testing set is kept out and used exclusively for reporting all downstream task results shown in the tables throughout the paper. An exception is made for the figures generated from the MLP-based downstream models, where we combine the training and validation sets for model training, as explained in Section~\ref{sec:dataset_config}.

\subsection{Sample Selection for External Validation}
\label{sec:external_val}

We use multiple external datasets for validation. 
For LAMOST LRS, we perform two types of comparisons: 
\begin{enumerate}
    \item \textbf{Radial Velocity:} Compared with GALAH DR4 \citep{2025PASA...42...51B}, where 49,905 stars are selected for which GALAH’s global RV fit succeeded (\texttt{FIT\_GLOBAL\_RV} = True), with \texttt{SNR\_PX\_CCD2} $\geq$ 20, LAMOST SNR$_g$ $\geq$ 20, and both GALAH \texttt{RV\_COMP\_1} and LAMOST RV having absolute values $\leq$ 999 km\,s$^{-1}$.
    \item \textbf{Iron Abundance:}
    \begin{itemize}
        \item DESI DR1 \citep{2025arXiv250314745D, 2025arXiv250514787K}: 119,335 stars are selected with LAMOST SNR$_g$ $\geq$ 30, DESI \texttt{SN\_B} $\geq$ 30, \texttt{SUCCESS} = True, and [Fe/H] $\geq -3.8$.
        \item GALAH DR4: 33,411 stars are retained with \texttt{FLAG\_FE\_H} = 0, \texttt{SNR\_PX\_CCD2} $\geq$ 30, LAMOST SNR$_g$ $\geq$ 30 in both surveys, and $T_{\rm eff}$ $\geq$ 4000~K.
    \end{itemize}
\end{enumerate}

For Gaia XP, the iron abundance validation uses:
\begin{enumerate}
    \item GALAH DR4: 33,411 stars satisfying \texttt{FLAG\_SP} = 0, \texttt{SNR\_PX\_CCD2} $\geq$ 50, and $T_{\rm eff}$ $\geq$ 4000~K.
    \item Gaia RVS catalog \citep{2024A&A...683L..11V}: 1,413 cross-matched stars are used.
\end{enumerate}

\subsection{Preprocessing}
\label{sec:preprocess}

For the LAMOST LRS spectrum, to focus on the most informative spectral features, we retain the 400–560 nm wavelength range, resulting in 1462 flux points per spectrum. The spectra are normalized before being input into the model; the normalization procedure is described in Appendix~\ref{sec: normalization}, where an iterative polynomial fitting algorithm robustly estimates the stellar continuum across both blue and red wavelength segments while suppressing absorption features and noise. For the Gaia XP spectra, broad-band color information may sometimes carry more discriminative power than individual spectral features. We therefore normalize each spectrum by its flux at 550 nm which lies near the center of the $V$-band.

\begin{table*}[ht]
\centering
\scriptsize
\caption{Comparison of Model Performance (standard deviation of the residuals $\sigma$ and coefficient of determination $R^2$) for different models evaluating on the held-out test datasets}
\label{tab:compare_updated}
\begin{tabular}{lccccccr}
\hline
\hline
\multicolumn{8}{c}{\textcolor{red}{\textbf{LRS Models}}} \\
\hline
Parameter & Raw Spectra & Pre-trained & CLIP & CLIP-r & CLIP-p & CLIP-pr & CLIP-split \\
 & $\sigma$ / $R^2$ & $\sigma$ / $R^2$ & $\sigma$ / $R^2$ & $\sigma$ / $R^2$ & $\sigma$ / $R^2$ & $\sigma$ / $R^2$ & $\sigma$ / $R^2$ \\
\hline
\multicolumn{8}{l}{\textit{Atmospheric Parameters}} \\
$\mathrm{[Fe/H]}$ & 0.070 / $-$0.882 & 0.066 / 0.939 & 0.058 / 0.949 &0.057/0.949 & 0.058 / 0.949 & 0.057 / 0.949 & \textbf{0.056} / \textbf{0.954} \\
$T_{\rm eff}$ (K) & 225.733 / 0.863 & 147.344 / 0.989 & 131.069 / \textbf{0.990} &137.360/\textbf{0.990} & 131.095 / \textbf{0.990} & \textbf{128.065} / \textbf{0.990} & 132.669 / \textbf{0.990} \\
$T_{\rm eff}$-sbi (maf) (K)& 106.903 / 0.979 & 94.942 / \textbf{0.990} & 96.577 / \textbf{0.990} &94.930/\textbf{0.990} & 95.004 / \textbf{0.990} & 95.346 / \textbf{0.990} & \textbf{93.047} / \textbf{0.990} \\
$T_{\rm eff}$-sbi (nsf) (K)& 76.986 / 0.982 & 84.991 / \textbf{0.991}$^\dag$ & 85.365 / 0.990 & 84.763/\textbf{0.991}& 85.065 / 0.990 & 84.101 / \textbf{0.991} & \textbf{82.309} / \textbf{0.991} \\
$\log g$ & 0.101 / 0.958 & 0.091 / 0.981 & 0.086 / 0.982 & 0.084/0.983& 0.086 / 0.982 & 0.085 / 0.983 & \textbf{0.079} / \textbf{0.985} \\
$\log g$-sbi (maf) & 0.063 / 0.967 & \textbf{0.062} / 0.981 & 0.064 / 0.982 & 0.065/0.983& 0.065 / 0.984 & 0.066 / 0.983 & 0.064 / \textbf{0.985} \\
\hline
\multicolumn{8}{l}{\textit{Elemental Abundances}} \\
$\mathrm{[\alpha/Fe]}$ & 0.023 / 0.872 & 0.021 / 0.906 & \textbf{0.020} / 0.912 &\textbf{0.020}/0.913 &\textbf{0.020} / 0.911 & \textbf{0.020} / \textbf{0.916} & \textbf{0.020} / 0.911 \\
$\mathrm{[C/Fe]}$ & 0.041 / 0.758 & 0.039 / 0.792 & \textbf{0.037} / 0.813 &\textbf{0.037}/0.812 &\textbf{0.037} / 0.812 & \textbf{0.037} / \textbf{0.814} & \textbf{0.037} / 0.813 \\
$\mathrm{[N/Fe]}$ & 0.054 / 0.598 & 0.052 / 0.642 & \textbf{0.049} / 0.664 &\textbf{0.049}/0.665 &\textbf{0.049} / 0.664 & \textbf{0.049} / \textbf{0.667} & \textbf{0.049} / \textbf{0.667} \\
$\mathrm{[Al/Fe]}$ & 0.049 / 0.691 & 0.048 / 0.711 & \textbf{0.046} / \textbf{0.741} &\textbf{0.046}/0.739 & \textbf{0.046} / 0.738 & \textbf{0.046} / 0.738 & \textbf{0.046} / 0.736 \\
$\mathrm{[Ca/Fe]}$ & 0.032 / 0.670 & 0.030 / 0.697 & \textbf{0.029} / 0.719 &\textbf{0.029}/0.720 & \textbf{0.029} / 0.721 & \textbf{0.029} / \textbf{0.723} & \textbf{0.029} / 0.714 \\
$\mathrm{[Mg/Fe]}$ & 0.031 / 0.866 & 0.032 / 0.871 & \textbf{0.031} / 0.882 &\textbf{0.031}/0.882 & \textbf{0.031} / 0.881 & \textbf{0.031} / \textbf{0.883} & \textbf{0.031} /0.880 \\
$\mathrm{[Si/Fe]}$ & 0.029 / 0.776 & 0.029 / 0.803 & \textbf{0.028} / \textbf{0.813} &\textbf{0.028}/\textbf{0.813} & \textbf{0.028} /  \textbf{0.813} & \textbf{0.028} / 0.812 & \textbf{0.028} / 0.812 \\
$\mathrm{[Ti/Fe]}$ & 0.061 / 0.492 & 0.058 / 0.532 & 0.056 / 0.550 &\textbf{0.055}/0.551 & 0.056 / 0.551 & \textbf{0.055} / \textbf{0.552} & 0.056 / 0.551 \\
$\mathrm{[Mn/Fe]}$ & 0.033 / 0.761 & 0.032 / 0.780 & \textbf{0.031} / 0.796 &\textbf{0.031}/\textbf{0.798} &\textbf{0.031} / 0.797 & \textbf{0.031} / \textbf{0.798} & \textbf{0.031} / 0.793 \\
$\mathrm{[Ni/Fe]}$ & 0.027 / 0.426 & 0.026 / 0.454 & \textbf{0.025} / \textbf{0.490} & \textbf{0.025}/0.486& \textbf{0.025} / 0.486 & \textbf{0.025} / 0.488 & \textbf{0.025} / 0.485 \\
$\mathrm{[O/Fe]}$ & 0.051 / 0.698 & 0.050 / 0.722 & 0.049 / 0.729 & 0.049/\textbf{0.730}& 0.049 / 0.728 & \textbf{0.048} / \textbf{0.730} & 0.049 / 0.729 \\
$\mathrm{[Cr/Fe]}$ & 0.081 / 0.177 & 0.076 / 0.225 & \textbf{0.074} / \textbf{0.242} &0.075/0.240 & 0.075 / 0.239 & 0.075 / 0.239 & 0.075 / 0.232 \\
\hline
\multicolumn{8}{l}{\textit{Asteroseismic Parameters}-sbi (maf)} \\
$\Delta\nu$ & \textbf{1.372}/0.901 & 1.705/0.958 & 1.491/0.841 &1.515/0.859 & 1.630/0.818 & 1.491/0.862 & 1.507/\textbf{0.963} \\
$\nu_{\rm max}$ & \textbf{20.470}/\textbf{0.676} & 23.597/0.330 & 22.590/0.171 &23.738/0.296 & 22.149/0.615 & 22.136/0.606 & 23.822/0.623 \\
Mass ($M_{\odot}$) & 0.095/0.518 & 0.094/0.570 & 0.086/\textbf{0.674} & 0.087/0.670 & \textbf{0.084}/0.669& 0.085/0.664 &0.089/0.658 \\
Radius ($R_{\odot}$) & \textbf{0.604}/0.873 & 0.708/0.859 & 0.738/0.853 &  0.728/0.853& 0.723/0.843 & 0.737/0.847 & 0.713/\textbf{0.880} \\
Age (Gyr) & 1.565/0.655 & 1.488/0.684 & 1.397/0.721 &1.347/0.744 & 1.347/\textbf{0.751} & 1.352/0.747 & \textbf{1.337}/0.723 \\
$\Delta\Pi$ & 28.078/0.891 & 25.703/0.904 & 21.471/0.923 & 21.814/0.920& 22.057/0.917 & \textbf{21.346}/0.924 & 22.713/\textbf{0.925} \\
\hline
\multicolumn{8}{l}{\textit{Other Parameters}} \\
$E(BP-RP)$ & 0.075 / $-$36.886 & 0.076 / 0.711 & 0.070 / 0.739 &0.070/0.740 & \textbf{0.069} / 0.742 & \textbf{0.069} / \textbf{0.743} & 0.072 / 0.741 \\
$v_{r}$ (km~s$^{-1}$) & 6.071 / 0.970 & 5.345 / 0.978 & 6.782 / 0.969 & 6.158/0.972&6.749 / 0.969 & 6.243 / 0.972 & \textbf{5.289} / \textbf{0.979} \\
$v_{r}$-sbi (maf) (km~s$^{-1}$)& \textbf{4.573} / 0.963 & 4.653 / \textbf{0.979} & 5.774 / 0.959 & 5.238/0.959& 5.786 / 0.960 & 5.270 / 0.961 & 4.581 / 0.978 \\
\hline
\multicolumn{8}{c}{\textcolor{blue}{\textbf{XP Models}}} \\
\hline
Parameter & Raw Spectra & Pre-trained & CLIP & CLIP-r & CLIP-p & CLIP-pr & CLIP-split\\
 & $\sigma$ / $R^2$ & $\sigma$ / $R^2$ & $\sigma$ / $R^2$ & $\sigma$ / $R^2$ & $\sigma$ / $R^2$ & $\sigma$ / $R^2$& $\sigma$ / $R^2$ \\
\hline
\multicolumn{8}{l}{\textit{Atmospheric Parameters}} \\
$\mathrm{[Fe/H]}$ & 0.469 / -0.389 & 0.126 / 0.884 & \textbf{0.111} / \textbf{0.900} & \textbf{0.111}/\textbf{0.900} & \textbf{0.111} / \textbf{0.900} & 0.112 / 0.899 & 0.113 / 0.894 \\
$T_{\rm eff}$ (K) & 220.258 / 0.965 & 199.458 / 0.969 & 172.722 / \textbf{0.974} &\textbf{169.602}/\textbf{0.974} & 172.638 / \textbf{0.974} & 171.811 / \textbf{0.974} & 170.696 / 0.973\\
$T_{\rm eff}$-sbi (maf) (K)& \dots / \dots$^\ddag$ & \dots / \dots & 173.804 / 0.968 & \dots/\dots& \dots / \dots & \dots / \dots  & \dots / \dots \\
$T_{\rm eff}$-sbi (nsf) (K)& 137.247 / 0.970 & 150.437 / 0.964 & 130.980 / 0.970 &132.889/0.971 & 130.689 / \textbf{0.972} & \textbf{129.708} / 0.970  & 131.251 / \textbf{0.972} \\
$\log g$ & 0.757 / 0.580 & 0.206 / 0.953 & 0.175 / 0.962 &0.173/0.962 & 0.173 / 0.962 & \textbf{0.171} / \textbf{0.963} & 0.174 / 0.961\\
$\log g$-sbi (maf) & 0.202 / 0.941 & 0.182 / 0.952 & 0.166 / \textbf{0.959} & 0.166/\textbf{0.959}& 0.165 / 0.958 & 0.167 / \textbf{0.959} & \textbf{0.164} / \textbf{0.959} \\
\hline
\multicolumn{8}{l}{\textit{Elemental Abundances}} \\
$\mathrm{[\alpha/Fe]}$ & 0.103 / $-$0.047 & 0.056 / 0.737 & 0.049 / 0.774 &\textbf{0.048}/\textbf{0.777} & 0.049 / 0.773 & 0.049 / 0.770 & 0.050 / 0.765 \\
$\mathrm{[C/Fe]}$ & 0.194 / 0.073 & 0.127 / 0.527 &  0.118 / 0.547  & 0.118/\textbf{0.553}& 0.118 / 0.550 & \textbf{0.117} / 0.549 & 0.118 / 0.551 \\
$\mathrm{[N/Fe]}$ & 0.115 / $-$4.040 & 0.077 / 0.643 & \textbf{0.072} / 0.673 & \textbf{0.072}/\textbf{0.676} & 0.073 / 0.672 & \textbf{0.072} / 0.674 & 0.073 / 0.669\\
\hline
\multicolumn{8}{l}{\textit{Other Parameters}} \\
$E(BP-RP)$ & 0.077 / 0.725 & 0.036 / 0.921 & 0.036 / 0.926 & \textbf{0.035}/0.927& 0.036 / 0.925 & \textbf{0.035} / 0.926 & \textbf{0.035} / \textbf{0.929} \\
\hline
\hline
Number of wins (best $\sigma$ or $R^2$) &0&0&19&24&15&29&19 \\
\hline
\hline
\end{tabular}
\begin{flushleft}
\textit{Note.} 
``CLIP'' for contrastive training-only model, ``CLIP-r'' for CLIP+reconstruction (LRS/XP) decoders, ``CLIP-p'' for CLIP+cross decoders, ``CLIP-pr'' for CLIP+all decoders, ``CLIP-split'' for CLIP+all decoders and an explicit separation of shared and non-shared embedding spaces. Most values are run with down-stream models of MLP but the ones with "-sbi" suffix are generated by the SBI models, where two kinds of SBI models, MAF and NSF models, are applied. Numbers in bold indicate the best performance (i.e., lowest $\sigma$ or highest $R^2$) for each parameter across all models. The last row reports the number of times each model achieves the best performance (i.e., lowest $\sigma$ or highest $R^2$) for any parameter, based on results from MLP-based downstream models only. Some outliers in prediction may dominate the overall $R^2$, occasionally leading to negative $R^2$ values. The numbers are reported as the average over 5 independent training runs, except for the results using SBI. For these, we report the single best-performing run among the five, selected based on a simulation-based calibration (SBC) test with a $p$-value threshold of at least 0.05, and prioritized by the lowest $\sigma$. $^\dag$Results marked exclude a failed NSF sampling case on one extreme spectrum. $^\ddag$Entries with dots indicate cases where all five runs failed the SBC test and no further tuning was performed. The same convention applies to other tables in this paper. We report the robust standard deviation of residuals ($\sigma$) using the Tukey Biweight Scale Estimator \citep{hoaglin1983understanding}, implementation available in \href{https://github.com/Xiaosheng-Zhao/SpecCLIP/blob/main/utils/robust_sigma.py}{\texttt{robust\_sigma.py}}, in all tables where $\sigma$ was used for internal model comparisons. For the plots involving external comparisons, we instead use \texttt{sigma\_clip} from \texttt{astropy} with 3$\sigma$ clipping, for ease of replication.
\end{flushleft}
\end{table*}

\begin{table}[ht]
\centering
\caption{Comparison of Cross-Modal Prediction Errors and Similarity Scores in the Projected Embeddings}
\label{tab:compare_similarity_pred_score}
\begin{tabular}{lrrr}
\hline
Model & XP → LRS MSE & LRS → XP MSE & Similarity \\
\hline
CLIP           & \dots                         & \dots                         & 0.7828                   \\
CLIP-r           & \dots                     & \dots      & 0.7783                   \\
CLIP-p           & 0.3932                     & 3.20 $\times 10^{-3}$      & 0.7820                   \\
CLIP-pr      & 0.3934                     & \textbf{3.15 $\times 10^{-3}$}      & 0.7778                   \\
CLIP-split          & \textbf{0.3929}                     & 3.48 $\times 10^{-3}$      & \textbf{0.7854}                   \\

\hline
\end{tabular}
\begin{flushleft}
\textit{Note.} 
Numbers in bold indicate the best performance (i.e., lowest MSE or highest similarity scores) across all models. Entries with dots indicate that the corresponding models are not applicable to the task.
\end{flushleft}
\end{table}

\section{Results}
\label{sec:results}

This section presents a comprehensive evaluation of SpecCLIP across multiple dimensions. We begin by comparing model variants, followed by parameter-estimation results for representative parameters using both LAMOST LRS and Gaia XP spectra. We end with demonstrations of spectral retrieval and cross-modal prediction.

\subsection{Model Comparison}
\label{sec:model_compare}

Table~\ref{tab:compare_updated} summarizes the overall performance of different models on held-out test datasets for parameter estimation. The pre-trained model on LAMOST LRS spectra generally outperforms the raw spectra, and importantly, CLIP-based models consistently improve performance for both LAMOST LRS and Gaia XP spectra, in most tasks where raw spectra or pre-trained (on LAMOST LRS or Gaia XP only) models alone were less effective. One notable exception is the radial velocity $v_r$ from LAMOST LRS, where most CLIP-based models perform worse than the LRS pre-trained model. This is understandable, as radial velocity is primarily determined by line features in LAMOST LRS spectra, and the alignment between LAMOST and Gaia -- taken at slightly different stellar epochs -- may introduce inconsistencies that degrade performance. Another exception is the \textit{Asteroseismic Parameters}-sbi task, where no clear differences are observed among models, possibly due to the small dataset size (3,029 stars). These results highlight the value of CLIP-based alignment. 

In particular, if checked more carefully, models with in-modal reconstruction decoders (e.g., CLIP-r and CLIP-pr compared with CLIP, and CLIP-pr compared with CLIP-p) generally show improved performance, as revealed by the ``number of wins'' in the table (last row), fairly accounting for the MLP-based downstream models only. This indicates enhanced informativeness of the learned representations for the estimation of downstream parameters. Overall, these results suggest that the inclusion of in-modal reconstruction decoders improves representation quality in most downstream applications.

We hypothesize that these performance gains come from the model’s ability to retain shared and modality-specific (non-shared) information. In the CLIP-pr variant, this is encouraged by jointly optimizing the contrastive loss, cross-modal prediction loss, and in-modal reconstruction loss. This architecture implicitly encourages the embeddings to retain complementary information from each modality.

To further explore this hypothesis, we introduce the CLIP-split model, which explicitly separates the projected embeddings into a 512-dimensional shared space and a 128-dimensional non-shared space. Despite using fewer parameters (see Appendix~\ref{sec:proj_decode}), CLIP-split performs competitively, particularly for core stellar parameters such as $T_{\rm eff}$, $\log g$, and [Fe/H]. It also recovers radial velocity performance to a level comparable with the LAMOST LRS pre-trained model, suggesting that the embedding split scheme retains more LRS-specific line features relevant to RV estimation.

Beyond parameter estimation, we also evaluated the models on two additional tasks using 50,000 paired spectra selected from the validation split of the datasets used for contrastive training with decoders.

\begin{itemize}
    \item \textbf{Similarity Score:} Measures how closely projected embeddings (or shared embeddings for CLIP-split) from different modalities align, which is crucial for cross-modal retrieval. Higher scores indicate more effective alignment.
    \item \textbf{Cross-Modal Prediction Score:} Evaluates the weighted (by measurement error \footnote{For comparison purposes (not exactly strictly), for LAMOST LRS, we propagate the inverse variance ($ivar$) of the flux measurements by multiplying by the square of the continuum fit ($C^2$), transforming $ivar$ to $ivar \cdot C^2$ for the normalized spectrum. For Gaia XP spectra, when normalizing the flux ($F$) by the flux at 550nm ($F_{550}$), the error ($\sigma_{F_N}$) of the normalized flux ($F_N = F / F_{550}$) is calculated using the standard error propagation for division: $\sigma_{F_N} = F_N \sqrt{(\sigma_F/F)^2 + (\sigma_{F_{550}}/F_{550})^2}$, where $\sigma_F$ and $\sigma_{F_{550}}$ are the respective flux errors.}) mean squared error (MSE) between predicted and ground-truth spectra in cross-modal translation (e.g., LRS $\rightarrow$ XP or XP $\rightarrow$ LRS).
\end{itemize}

These results are summarized in Table~\ref{tab:compare_similarity_pred_score}. We find that models with both reconstruction and prediction decoders (CLIP-pr) yield improved performance on LRS $\rightarrow$ XP prediction, but slightly degrade the similarity score, an expected trade-off when the embeddings are trained to retain both shared and non-shared information. Nevertheless, their similarity scores remain significantly higher than the baseline similarity (0.0533) obtained from comparing the embeddings between modality-specific pre-trained models.

CLIP-split achieves the highest similarity score overall, even surpassing the baseline CLIP model, possibly aided by its lower embedding dimensionality, which tends to produce higher cosine similarities. It also delivers the best performance on the prediction of the LRS spectrum of XP $\rightarrow$ LRS, demonstrating the robustness of the model in all modalities.

In summary, models with prediction and reconstruction decoders (CLIP-pr and CLIP-split) offer the best overall performance by balancing parameter-estimation accuracy, cross-modal predictability, and embedding similarity. 

\begin{figure*}
\begin{center}
\includegraphics[scale=0.36,angle=0]{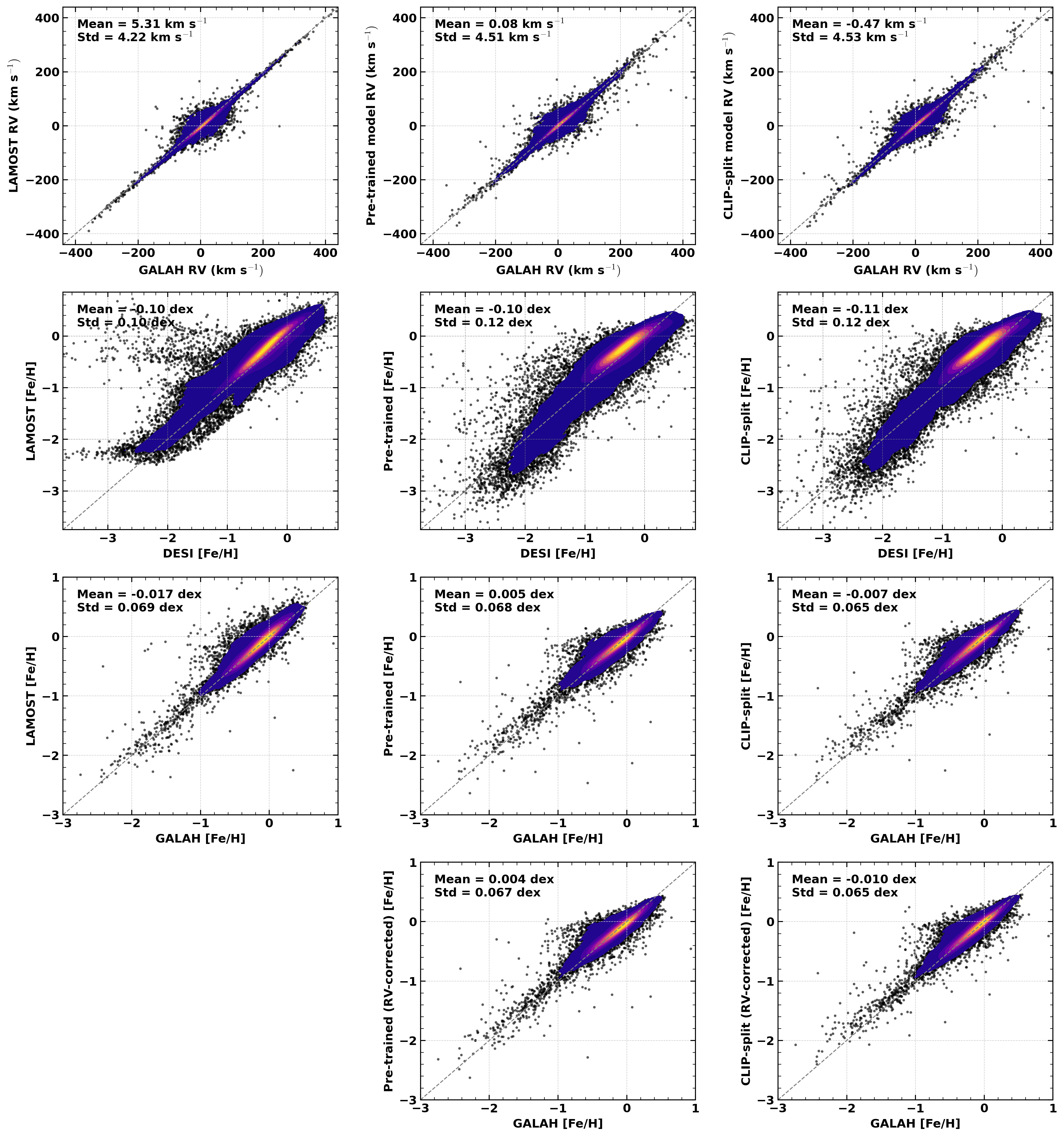}
\caption{Comparison between the LAMOST catalog and SpecCLIP models (including the pre-trained LRS model and the LRS branch of the CLIP-split model). From top to bottom: The radial velocity (RV) comparison  as a function of the GALAH labels; [Fe/H] comparison as a function of the DESI labels; [Fe/H] comparison as a function of the GALAH labels; and [Fe/H] comparison as a function of the GALAH labels with input spectra shifted to the rest frame using the predicted RVs from the corresponding models in the top row. For RV, which is inferred using the SBI downstream model, the pre-trained LRS model and CLIP-split model have slightly larger scatter but smaller bias, compared with the LAMOST catalog; For [Fe/H], inferred using MLP downstream models (as with all other figures), the pre-trained LRS model and CLIP-split model gives either smaller scatter over the metal-poor region (referring to DESI labels) or overall smaller scatter and bias (referring to GALAH labels). The RV-corrected spectra result in similar [Fe/H] prediction performance, suggesting that the trained MLP models are relatively robust to modest Doppler shifts in the LAMOST LRS spectra. The dashed lines are the one-to-one lines.  The numbers in the upper left of each panel are the mean offsets and standard deviation of the residuals (y-axis minus x-axis).}
\label{fig:results_lrs_feh}
\end{center}
\end{figure*}

\subsection{Parameter Estimation}
\label{sec:estimation}

\begin{figure*}
\begin{center}
\includegraphics[scale=0.38,angle=0]
{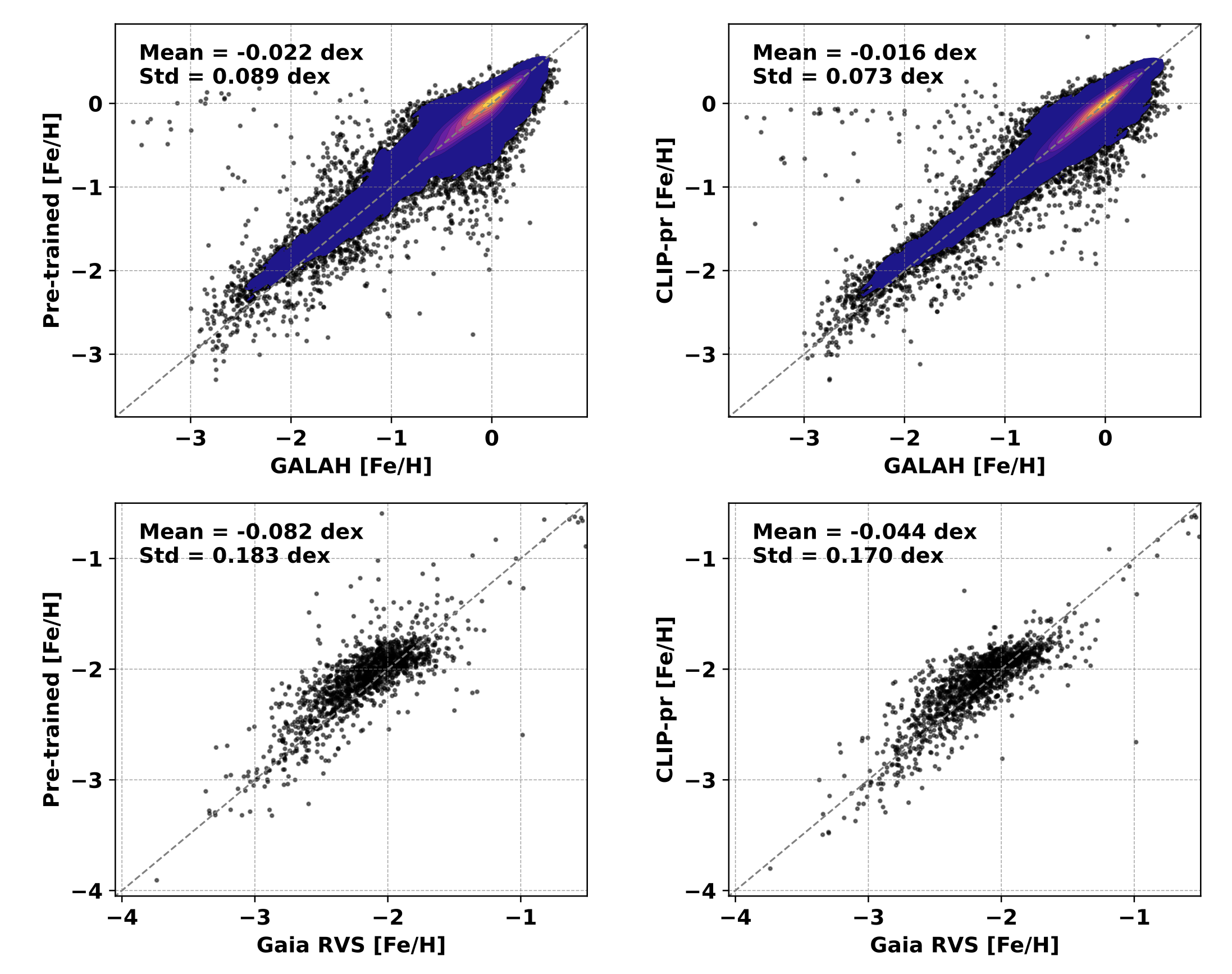}
\caption{Comparison of [Fe/H] estimates from SpecCLIP (pre-trained XP model and XP branch of the CLIP-pr model) with reference labels from GALAH (top) and Gaia RVS (bottom). Both models correlate well with reference labels, with the CLIP-pr model yielding lower scatter and bias. The dashed lines are the one-to-one lines.  The numbers is the upper left of each panel are the mean offsets and standard deviation of the residuals.}
\label{fig:results_xp_FeH}
\end{center}
\end{figure*}

\subsubsection{LAMOST LRS}
\label{sec:lrs}

Figure~\ref{fig:results_lrs_feh} presents results for radial velocity and iron abundance estimation\footnote{The reported \texttt{std} values for all plots are calculated using \texttt{sigma\_clip} from \texttt{astropy} with 3$\sigma$ clipping.}. For radial velocity, we compare predictions from the LRS pre-trained model and the CLIP-split model with the official LAMOST stellar parameter catalog, using GALAH DR4 -- which is not included in the training set -- as an external benchmark. While our models produce slightly larger standard deviations (4.51 and 4.53 km~s$^{-1}$ compared to 4.22 km~s$^{-1}$ from the official LAMOST pipeline), they exhibit significantly smaller biases, producing values that are closer to the true measurements.

From a computational perspective, our LRS-based RV inference is highly efficient: In an environment with 1 core (Intel® Xeon® Gold 6248 @ 2.50GHz) and 1 V100 GPU, inference takes $\sim$5ms per spectrum using SBI, and 1ms per spectrum using an MLP. However, as seen in Table~\ref{tab:compare_updated}, SBI models provide better precision (lower $\sigma$), while MLPs offer comparable or better accuracy ($R^2$). 

For iron abundance [Fe/H], we benchmark our models against the DESI DR1 and GALAH DR4. Compared to DESI, the CLIP-split model demonstrates significantly better performance than the LAMOST official pipeline. The plateau in [Fe/H] around $-2.5$, which arises from the lack of metal-poor stars in the stellar library used by the LAMOST pipeline, is effectively addressed by our models, although the overall scatter remains comparable. When benchmarked against GALAH, the CLIP-split model achieves both lower bias and lower scatter, highlighting its competitive performance relative to physically motivated pipelines. We further investigated the impact of rest-frame correction by applying the predicted RVs to shift the spectra before feeding them into the [Fe/H] prediction models. This additional step yields predictions that are broadly consistent with those from uncorrected inputs, indicating that correcting for radial velocity is not critical—at least for the resolution and wavelength coverage of LAMOST LRS. The insensitivity of the MLP-based [Fe/H] predictions to small redshifts implies that the model has implicitly learned to accommodate these variations.

Although our main pipeline estimates one parameter per MLP model (multi-variate MLP is also straightforward, though we did not explore it in this paper), we also experimented with SBI variants that estimate either one or all parameters jointly. While overall performance was similar, joint estimation, especially with SBI, better captures parameter degeneracies. An example of inferred chemical abundances with SBI is shown in Appendix~\ref{sec:pred_more}, Figure~\ref{fig:results_lrs_abundance}.

In general, our method compares favorably with previous work \citep{2019ApJS..245...34X, 2023MNRAS.521.6354L, 2023ApJS..266...40W, 2025arXiv250602763Z,2025ApJS..278...41Z}. Compared with the other data-driven methods listed, our approach generally requires fewer labeled training samples (typically$<$90,000) and minimal hyperparameter tuning, as we adopt a unified architecture for all downstream models. Relative to physics-driven methods such as template fitting or forward modeling, our model does not require synthetic spectral templates or explicit physical modeling at inference time. This design, combined with diverse training data, enables applicability across a wide range of stellar types and delivers fast predictions once trained. Although our method avoids physical modeling during inference, its effectiveness still depends on high-quality labels, all of which are ultimately derived from physics-based modeling approaches. These characteristics make our method both efficient and broadly applicable in practice. However, direct comparison with previous literature remains challenging due to differences in the composition of the test set. For example, our iron abundance test set extends to [Fe/H]~$\sim -4$, increasing the difficulty of achieving a general low scatter or a high $R^2$.

\subsubsection{Gaia XP}
\label{sec:xp}

Figure~\ref{fig:results_xp_FeH} shows [Fe/H] predictions from the Gaia XP model (both pre-trained and CLIP-split models) compared with ground truth from the GALAH DR4 and Gaia RVS catalog. Our predictions are consistent across the entire iron abundance range, including the metal-poor regime down to about [Fe/H] $= -3.5$ or even $-4.0$, and outperform previous machine learning methods \citep{2023ApJS..267....8A}, particularly at our ability to extend to low iron abundances.
The overall scatter is below 0.08~dex for stars with [Fe/H]~$> -2.0$, and below 0.18~dex for stars with [Fe/H]~$< -2.0$.
This performance surpasses that achieved by traditional low-resolution spectroscopy.

\begin{figure}
\begin{center}
\includegraphics[width=\columnwidth]{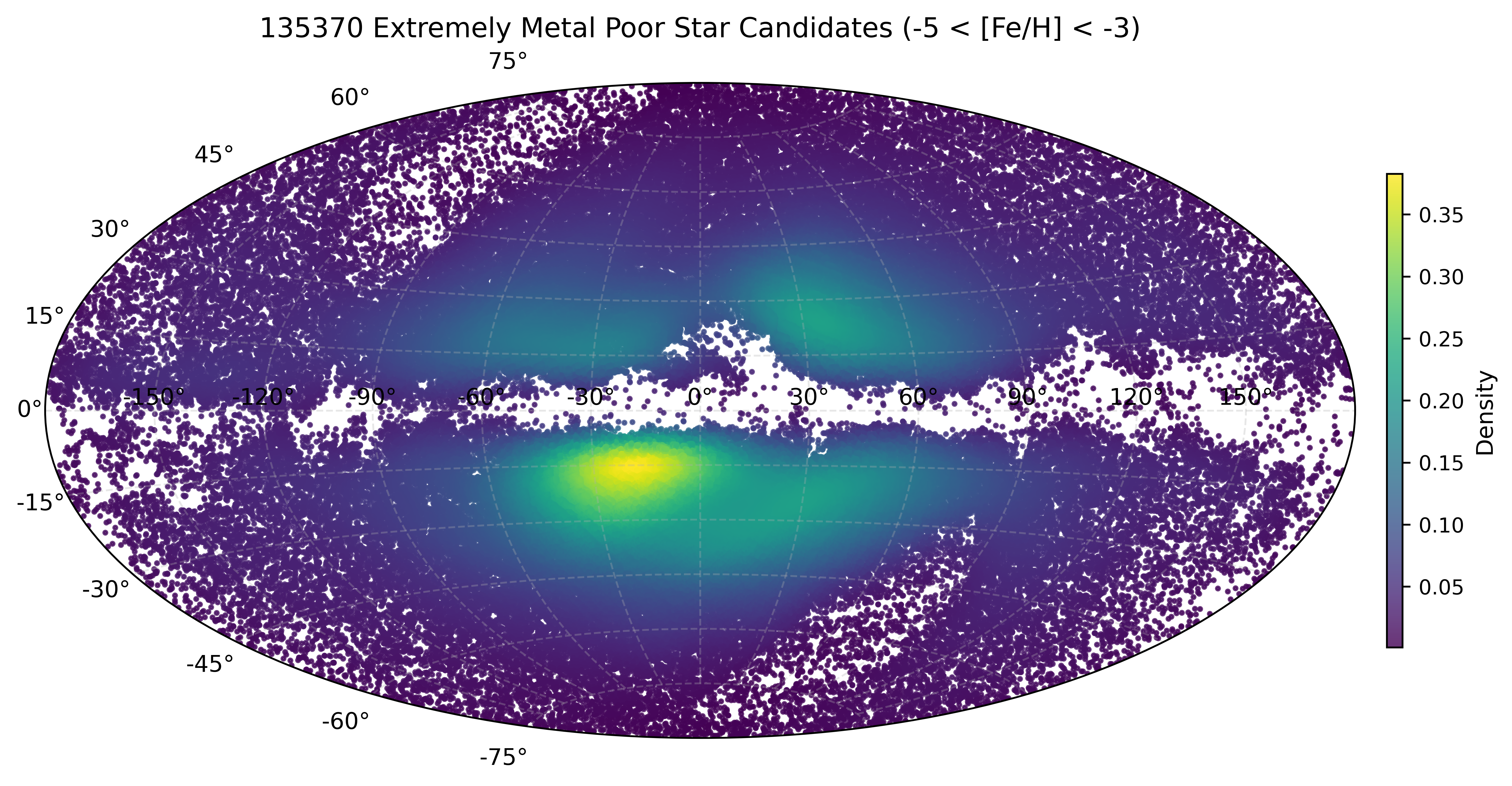}
\caption{Spatial density distributions of extremely metal-poor stars ($-5<{\rm [Fe/H]}<-3$) derived from SpecCLIP (CLIP-pr model) in Galactic coordinates, showing a clear ``metal-poor old heart'' of our Galaxy.}
\label{fig:results_xp_distribution}
\end{center}
\end{figure}

Figure~\ref{fig:results_xp_distribution} highlights 135,370 extremely metal-poor (EMP) star candidates with iron abundances in the range $-5 < \mathrm{[Fe/H]} < -3$, identified by our XP branch of the CLIP-pr model. These stars exhibit a pronounced concentration toward the Galactic center, reminiscent of the ``metal-poor heart of the Galaxy" reported by \citet{2022ApJ...941...45R}, but now extending to significantly lower iron abundances than previously observed. A dedicated follow-up study based on this sample is currently underway, aiming to shed light on the earliest phases of the Milky Way’s chemical and structural evolution.

Performance metrics in various XP models are shown in Table~\ref{tab:compare_updated}. Again, CLIP-based models (CLIP, CLIP-pr, CLIP-split) are competitive compared to earlier approaches, including \citet{2024ApJ...974..192H} and \citet{2024ApJS..272....2L}. Notably, our test sets span a wide parameter range.  For example, [Fe/H] extends down to around $-4.0$ and $T_{\rm eff}$ extends up to 13,500~K, further validating the robustness of our models.

\begin{figure*}
\begin{nolinenumbers}
    \begin{tikzpicture}
    \centering
        \node[anchor=south,inner sep=0] (first) at (0,0) {
            \includegraphics[scale=0.225]{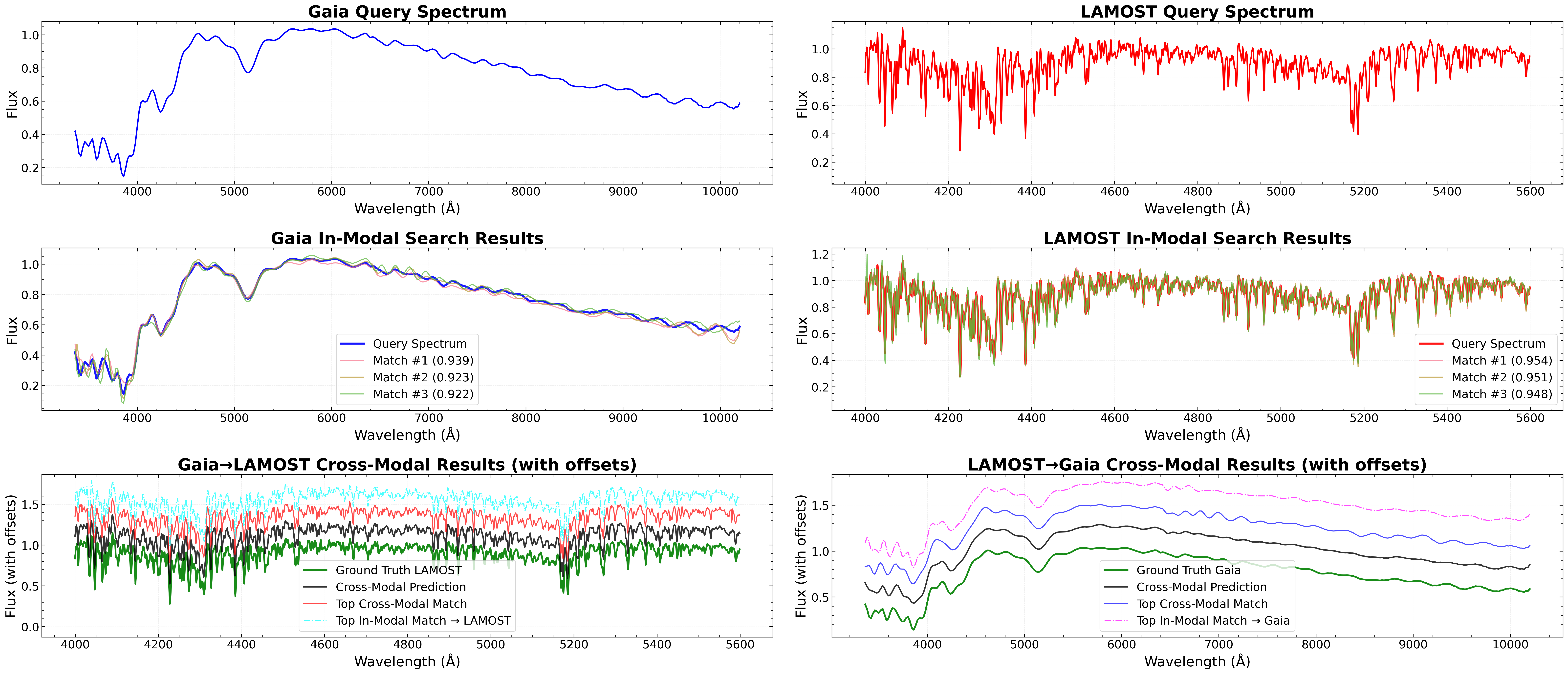}
        };

        \node[anchor=north,inner sep=0] (second) at ([yshift=-.5cm]first.south) {
            \includegraphics[scale=0.225]{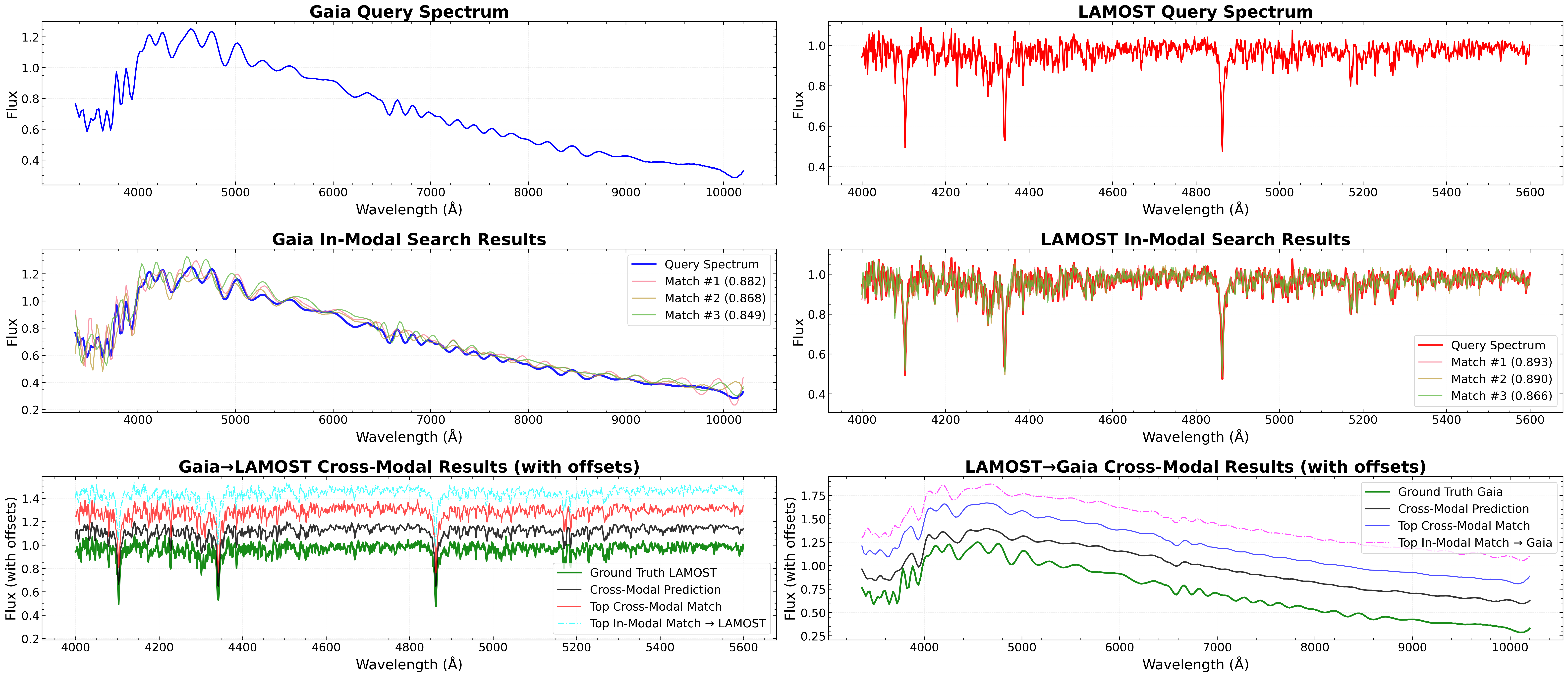}
        };
    \end{tikzpicture}
    \end{nolinenumbers}
    \caption{Two examples of in-modal retrieval, cross-modal retrieval, cross-modal prediction, and the LAMOST (Gaia) spectra corresponding to Gaia (LAMOST) in-modal retrieval. The similarity scores are defined in the projected embedding space.}
    \label{fig:results_clip_cross_modality_search}
\end{figure*}

\begin{table*}[ht]
\centering
\scriptsize
\caption{Comparison of Model Performance (standard deviation of the residuals $\sigma$ and coefficient of determination $R^2$) for pre-trained XP models with different embedding dimensions (256, 343, 512, 768), where 768 is adopted in this paper}
\label{tab:compare_updated_dense_xp_embed}
\begin{tabular}{lccccc}
\hline
\multicolumn{6}{c}{\textcolor{blue}{\textbf{XP Models}}} \\
\hline
Parameter & Raw Spectra & 256 & 343 & 512 & 768 \\
 & $\sigma$ / $R^2$ & $\sigma$ / $R^2$ & $\sigma$ / $R^2$ & $\sigma$ / $R^2$& $\sigma$ / $R^2$  \\
\hline
\multicolumn{6}{l}{\textit{Atmospheric Parameters}} \\
$\mathrm{[Fe/H]}$ & 0.469 / $-$0.389 & 0.128 / 0.881& 0.127 / 0.882& 0.129 / 0.879  & \textbf{0.126} / \textbf{0.884}  \\
$T_{\rm eff}$ (K) & 220.258 / 0.965 & 201.126 / 0.967 & 203.142 / 0.967& \textbf{196.401} / 0.968& 199.458 / \textbf{0.969}\\
$\log g$ & 0.757 / 0.580 & 0.207 / 0.952 & 0.208 / 0.951& \textbf{0.197} / \textbf{0.955} & 0.206 / 0.953\\
\hline
\multicolumn{6}{l}{\textit{Elemental Abundances}} \\
$\mathrm{[\alpha/Fe]}$ & 0.103 / $-$0.047 & 0.057 / 0.732 & 0.057 / 0.731 & \textbf{0.056} / \textbf{0.738}& \textbf{0.056} / 0.737\\
$\mathrm{[C/Fe]}$ & 0.194 / 0.073 & 0.129 / \textbf{0.527} & 0.128 / 0.525& 0.129 / 0.526& \textbf{0.127} / \textbf{0.527} \\
$\mathrm{[N/Fe]}$ & 0.115 / $-$4.040 & 0.079 / 0.631 & 0.080 / 0.628 & \textbf{0.077} / 0.641& \textbf{0.077} / \textbf{0.643} \\
\hline
\multicolumn{6}{l}{\textit{Other Parameters}} \\
$E(BP-RP)$ & 0.077 / 0.725 & 0.038 / 0.913 & 0.039 / 0.915 & \textbf{0.036} / 0.915& \textbf{0.036} / \textbf{0.921}\\
\hline
\end{tabular}
\begin{flushleft}
\textit{Note.} 
Numbers in bold indicate the best performance (i.e., lowest $\sigma$ or highest $R^2$) for each parameter across all models.
\end{flushleft}
\end{table*}

\subsection{Spectral Retrieval and Prediction}
\label{sec:retrieval_prediction}

\subsubsection{Spectral Retrieval}
\label{sec:search}

Beyond parameter estimation, SpecCLIP also enables retrieval of similar spectra within the learned embedding space, both within a single modality and across different modalities.

Figure~\ref{fig:results_clip_cross_modality_search} shows examples of in-modal and cross-modal retrievals given a specific query spectrum. Retrieval is based on cosine similarity in the (projected) embedding space, using either the full or shared embeddings (for CLIP-split). In this figure, the search is performed using the CLIP-pr model on a test set of 82,057 spectra, with the query spectrum excluded from the database. In both LAMOST LRS and Gaia XP cases, the retrieved spectra closely resemble the query spectra, indicating that the model has learned well-aligned representations across modalities.

In practice, additional strategies can be used to retrieve spectra using auxiliary catalog links. For example, given a query LAMOST spectrum and a LAMOST-to-Gaia cross-match library, one could first retrieve the top LAMOST matches (in-modal) and then fetch their Gaia counterparts via the library. Alternatively, if the paired Gaia spectrum of the query is known, one could retrieve similar Gaia spectra directly (in-modal), or indirectly by performing a Gaia-to-LAMOST retrieval followed by a database lookup to obtain the corresponding Gaia spectra. While Figure~\ref{fig:results_clip_cross_modality_search} presents only two retrieval use cases, assuming that we know only the information of the query spectrum itself, not its paired other-modal spectrum.
The other more elaborate approaches are straightforward extensions.

These capabilities suggest promising applications in data mining and search-based discovery. For instance, starting from a LAMOST spectrum of a rare type of star, one could search for spectrally similar candidates in the Gaia XP database of over two billion stars. Such functionality could significantly enhance large-scale searches for rare or unusual stellar types.

\subsubsection{Spectral Prediction}
\label{sec:pred}

SpecCLIP’s cross-modal decoders also support spectral translation, that is, predicting the spectrum in one modality from a spectrum in another. These decoders operate directly on the projected embeddings (shared embeddings in the case of CLIP-split), using learned mappings between modalities.

We find that for the majority of the test dataset, the model performs well in both directions (LRS $\rightarrow$ XP and XP $\rightarrow$ LRS), indicating that it effectively learns the mapping between these two spectroscopic modalities. Examples of such predictions are shown in Figure~\ref{fig:results_clip_cross_modality_search} as black curves.

However, for a subset of sources—particularly in the LRS $\rightarrow$ XP direction—prediction quality deteriorates. Apart from the potential effect of extinction, which may be poorly-captured by the trained model. This discrepancy may suggest that the source does not follow the behavior of a typical single star. For instance, it could be an unresolved binary or an otherwise anomalous object. These cases highlight an exciting future direction, using the cross-modal prediction error as a basis for anomaly detection.

We leave a more systematic exploration of anomaly detection and rare-object identification to future work.

\section{Discussion}
\label{sec:discussion}

\subsection{SBI Performance with NSF and MAF}
\label{sec:discuss_metalicity}

In Table~\ref{tab:compare_updated}, we present the SBI results for several parameters ($T_{\rm eff}$, $\log g$, $v_{r}$, and asteroseismic parameters) that demonstrate the benefits of adopting SBI. For parameters showing relatively clear differences between NSF and MAF, we include results from both; for those exhibiting visually biased predictions in one of the two SBI methods, we conservatively report only the other. For parameters modeled by both methods (e.g., $T_{\rm eff}$), we observe better performance with NSF, likely due to its higher flexibility in representing non-linear conditional dependencies between spectra and stellar parameters.

In our additional experiments, we observed that applying NSF to projected embeddings led to degraded performance in the low-iron abundance regime ($\mathrm{[Fe/H]} \lesssim -1.5$). Similarly, LAMOST LRS radial velocity predictions exhibited a systematic underestimation of the absolute value at large radial velocities; a similar issue is widely discussed in machine learning-based estimators \citep{2024arXiv241205806T}. Interestingly, replacing NSF with MAF eliminated these issues.

These results suggest that NSF is more expressive than MAF, this increased expressiveness can adversely affect performance in our applications when modeling distributions near parameter-space boundaries. A likely explanation is that the spline transformations used in NSF, while highly flexible within their domain, become problematic in sparsely sampled boundary regions. Specifically, NSF employs rational–quadratic splines over a finite interval (default: \(\pm 3\sigma\) in standardized space), with linear tail transformations beyond this range.

In astrophysical parameter spaces where extreme values (\([\mathrm{Fe/H}] \lesssim  -1.5\), \(|v_{\mathrm{rad}}| \gtrsim 150\,\mathrm{km\,s^{-1}})\) represent several percent of the population, these linear tails systematically mis-model the true distribution shape. Furthermore, the flexibility of spline binning can lead to overfitting in regions with sparse training coverage, where bin allocation becomes poorly constrained.

In contrast, MAF's simpler affine transformations naturally extend over the full unbounded support of the distribution with stable linear extrapolation, providing more robust behavior at distribution boundaries despite lower overall expressiveness. The contrasting performance between NSF and MAF highlights the importance of matching normalizing flow architecture to the characteristics of astrophysical parameter distributions, and motivates further investigation of boundary-aware flow designs for simulation-based inference in astronomy.

\subsection{Compression vs. Feature Learning}
\label{sec:discuss_embed}

A common assumption is that compression improves downstream performance. However, our Gaia XP foundation model shows a more nuanced behavior. In Table~\ref{tab:compare_updated_dense_xp_embed}, we vary the embedding dimension and find that reducing it from 343 (the original XP spectrum length) to 256 leads to comparable performance. In contrast, using embedding dimensions equal to or greater than the input size yields better results.

This suggests that effective feature learning. rather than compression alone, is key to high downstream performance in this case. Mapping 343 flux points into a higher-dimensional latent space (e.g., 768) may allow the model to capture richer nonlinear correlations and disentangle latent physical factors (e.g., $T_{\rm eff}$, $\log g$, [Fe/H], extinction), forming expressive embeddings for downstream prediction. We note, however, that increasing the embedding dimension also increases the number of trainable parameters in the downstream MLP, which may partly explain the improved performance. These interpretations remain tentative and will be further examined in future work.

\subsection{Interpretability of Parameter Estimation}
\label{sec:interpretability}

One potential concern is that the models may rely on spurious correlations in the data to estimate parameters. While forward models like Payne and DD-Payne allow straightforward inspection of such behavior, our neural networks do not offer easy interpretability.

Recent techniques, such as sparse autoencoders \citep{2023arXiv230908600C}, could improve the explainability of the model in future work. However, our model achieves high precision and accuracy on held-out datasets, suggesting that it is indeed learning physically meaningful representations. Further testing with carefully designed datasets will be necessary to validate this assumption and improve model interpretability.

\subsection{SBI vs. MLP}
\label{sec: sbi_vs_mlp}

Although we report MLP results for most tasks, we observe that SBI tends to outperform MLP in terms of uncertainty (as measured by $\sigma$), while MLP yields comparable or better accuracy (as measured by $R^2$), as shown in Table~\ref{tab:compare_updated}.

This discrepancy may arise from their differing training objectives;  MLPs minimize the MSE, which aligns closely with $R^2$, whereas SBI focuses on modeling the entire posterior distribution. In this work, we estimate the posterior median from SBI (rather than the mean), which is more robust to outliers. However, we note that differences in the adopted hyperparameters (see Appendix~\ref{sec:hyper_summary}) may also contribute to the discrepancy

Thus, for applications where uncertainty quantification is essential, SBI is the preferred choice. For fast and accurate point estimates, MLP is more suitable. In particular, our SBI models use only $\sim$0.1 million trainable parameters, compared to $\sim$1.4 million for the MLP models.

\subsection{Training Sample Size and Dataset Configuration}
\label{sec:dataset_config}

We observe a positive correlation between the size of the training set and the performance of the model in the downstream tasks. Although our benchmark experiments use $\sim$100,000 stars per parameter, we note that performance may not have plateaued, indicating room for improvement with additional data.

Nevertheless, even with this moderate sample size, our models match or outperform state-of-the-art results in the literature (Section~\ref{sec:estimation}). To maximize data usage, we report metrics based on held-out test sets, while for plots generated using the downstream models, we combine the training and validation sets for training, with early stopping based on performance on the test set.

Future work may adopt more advanced techniques such as $k$-fold cross-validation, which would allow iterative use of the entire dataset and further improve model reliability and performance.

\subsection{Is Transformer Overkill?}
\label{sec:transformer_vs_oae}

In Table~\ref{tab:compare_updated_dense_sbi_vs_mlp}, we compare MT and OAE  with equivalent numbers of trainable parameters and identical training epochs. For LAMOST LRS spectra, MT outperforms OAE. However, for Gaia XP spectra, OAE performs better.

These results suggest that transformers may be more advantageous for longer spectra (e.g., LRS with 1462 flux points), while offering limited benefits, or even unnecessary complexity, for shorter inputs like Gaia XP spectra (343 points). Another possible explanation is the difference in optimal training epochs for different architectures, as discussed in \citet{2025arXiv250318617R} and \citet{2025ApJ...980...66R}. This warrants further exploration to better match model capacity to data complexity.

\subsection{Fine-tuning Loss Weights}
\label{sec:weights}

Not all non-shared information in the spectra is necessarily beneficial—some components, such as noise, may even hinder downstream performance. Moreover, the importance of shared versus non-shared information can vary across tasks; some may depend more on modality-specific features, while others benefit primarily from shared representations. Therefore, fine-tuning the weight terms in Equation~\ref{eq:loss}, particularly the loss weight of reconstruction, may be a promising direction for future work.

In this work, we fix both weights to 1, applying equal weighting to the reconstruction and cross-modal prediction losses. This choice shows that models incorporating reconstruction loss already perform competitively without explicit weight adjustment, as evidenced by the number of highest performing metrics (i.e. “wins”) in Table~\ref{tab:compare_updated} (last row)—e.g., CLIP-r vs. CLIP, and CLIP-pr/CLIP-split vs. CLIP-p. However, the magnitude of performance gains remains modest and somewhat task-dependent. Reweighting the losses may further increase the total number of improvements or yield more substantial gains on specific parameters. However, the latter—optimizing for specific tasks—might come at the cost of generality, which is contrary to the fundamental goal of building a model that performs robustly across diverse downstream tasks.

\begin{table}[ht]
\centering
\scriptsize
\caption{Comparison of Model Performance (standard deviation of the residuals $\sigma$ and coefficient of determination $R^2$) between masked transformer (MT) and MLP-based ordinary auto-encoder (OAE)}
\label{tab:compare_updated_dense_sbi_vs_mlp}
\begin{tabular}{lccc}
\hline
\multicolumn{4}{c}{\textcolor{red}{\textbf{LRS Models}}} \\
\hline
Parameter & Raw Spectra & MT & OAE  \\
 & $\sigma$ / $R^2$ & $\sigma$ / $R^2$ & $\sigma$ / $R^2$ \\
\hline
\multicolumn{4}{l}{\textit{Atmospheric Parameters}} \\
$\mathrm{[Fe/H]}$ & 0.070 / $-$0.882 & \textbf{0.066} / \textbf{0.939}& 0.070 / 0.905 \\
$T_{\rm eff}$ (K)& 225.733 / 0.863 & \textbf{147.344} / \textbf{0.989} & 181.777 / 0.975 \\
$\log g$ & 0.101 / 0.958 & 0.091 / \textbf{0.981} & \textbf{0.084} / 0.973  \\
\hline
\multicolumn{4}{l}{\textit{Elemental Abundances}} \\
$\mathrm{[\alpha/Fe]}$ & 0.023 / 0.872 & 0.021 / \textbf{0.906} & \textbf{0.020} / 0.904\\
$\mathrm{[C/Fe]}$ & 0.041 / 0.758 & \textbf{0.039} / \textbf{0.792} & \textbf{0.039} / 0.776 \\
$\mathrm{[N/Fe]}$ & 0.054 / 0.598 & \textbf{0.052} / \textbf{0.642} & 0.053 / 0.624 \\
$\mathrm{[Al/Fe]}$ & 0.049 / 0.691 & \textbf{0.048} / \textbf{0.711} & 0.049 / 0.693\\
$\mathrm{[Ca/Fe]}$ & 0.032 / 0.670 & \textbf{0.030} / \textbf{0.697} & 0.031 / 0.688\\
$\mathrm{[Mg/Fe]}$ & 0.031 / 0.866 & 0.032 / 0.871 & \textbf{0.030} / \textbf{0.873}\\
$\mathrm{[Si/Fe]}$ & \textbf{0.029} / 0.776 & \textbf{0.029} / \textbf{0.803} & \textbf{0.029} / 0.793\\
$\mathrm{[Ti/Fe]}$ & 0.061 / 0.492 & \textbf{0.058} / \textbf{0.532} & 0.059 / 0.507 \\
$\mathrm{[Mn/Fe]}$ & 0.033 / 0.761 & \textbf{0.032} / \textbf{0.780} & 0.033 / 0.758 \\
$\mathrm{[Ni/Fe]}$ & 0.027 / 0.426 & \textbf{0.026} / \textbf{0.454} & 0.027 / 0.445\\
$\mathrm{[O/Fe]}$ & 0.051 / 0.698 & \textbf{0.050} / \textbf{0.722} & 0.051 / 0.704\\
$\mathrm{[Cr/Fe]}$ & 0.081 / 0.177 & \textbf{0.076} / \textbf{0.225} & 0.079 / 0.200\\
\hline
\multicolumn{4}{l}{\textit{Other Parameters}} \\
$E(BP-RP)$ & \textbf{0.076} / $-$23.199 & \textbf{0.076} / \textbf{0.711} & \textbf{0.076} / 0.681\\
$v_{r}$ (km~s$^{-1}$) & 6.418 / 0.942 & \textbf{5.345} / \textbf{0.978} & 5.938/0.966\\
\hline
\multicolumn{4}{c}{\textcolor{blue}{\textbf{XP Models}}} \\
\hline
Parameter & Raw Spectra & MT & OAE \\
 & $\sigma$ / $R^2$ & $\sigma$ / $R^2$ & $\sigma$ / $R^2$ \\
\hline
\multicolumn{4}{l}{\textit{Atmospheric Parameters}} \\
$\mathrm{[Fe/H]}$ & 0.469 / $-$0.389 & 0.137 / 0.867& \textbf{0.126} / \textbf{0.884}  \\
$T_{\rm eff}$ (K)& 220.258 / 0.965 & 215.299 / 0.965 & \textbf{199.458} / \textbf{0.969}\\
$\log g$ & 0.757 / 0.580 & \textbf{0.206} / \textbf{0.953} & \textbf{0.206} / \textbf{0.953}\\
\hline
\multicolumn{4}{l}{\textit{Elemental Abundances}} \\
$[\alpha/Fe]$ & 0.103 / $-$0.047 & 0.059 / 0.713 & \textbf{0.056} / \textbf{0.737} \\
$\mathrm{[C/Fe]}$ & 0.194 / 0.073 & 0.132 / 0.498 & \textbf{0.127} / \textbf{0.527} \\
$\mathrm{[N/Fe]}$ & 0.115 / $-$4.040 & 0.079 / 0.615 & \textbf{0.077} / \textbf{0.643} \\
\hline
\multicolumn{4}{l}{\textit{Other Parameters}} \\
$E(BP-RP)$ & 0.077 / 0.725 & 0.041 / 0.913 & \textbf{0.036} / \textbf{0.921}\\\\
\hline
\end{tabular}
\begin{flushleft}
\textit{Note.} 
Numbers in bold indicate the best performance (i.e., lowest $\sigma$ or highest $R^2$) for each parameter across all models.
\end{flushleft}
\end{table}

\vskip 1cm
\section{Summary}
\label{sec:summary}

In this work, we develop a foundation model framework for stellar spectra that enables strong and efficient performance across multiple downstream tasks. Our approach integrates separate pre-trained models, each trained on a distinct spectroscopic modality (LAMOST LRS or Gaia XP), and aligns them using CLIP-style contrastive learning. To further enhance the information capacity of the embeddings, we introduce decoder modules that increase the mutual information between the embeddings and input spectra and enable \textit{translation} (prediction) between different spectral types.

Our main findings are summarized below:

\begin{itemize}
    \item The pre-trained foundation models for both spectral modalities demonstrate strong performance with a relatively small number of labeled examples (i.e., few-shot learning). Using $\sim$100,000 stars with high-quality labels, they achieve competitive parameter inference performance across a range of stellar parameters. Comparisons with the LAMOST official release and high-resolution reference catalogs (e.g., GALAH, and APOGEE) confirm the accuracy and reliability of our method.
    
    \item Performance is further improved by contrastive alignment and the addition of decoders, which increase the robustness and expressiveness of the learned embeddings. These enhancements are especially beneficial for parameter estimation and spectral prediction.
    
    \item We explore the use of SBI as an alternative to MLPs for downstream parameter estimation. SBI provides improved uncertainty modeling and higher precision for certain parameters, albeit at a higher inference cost and a lower model capacity.
    
    \item Our models support both in-modal and cross-modal spectrum retrieval, as well as spectrum-to-spectrum prediction across modalities. High similarity and prediction scores demonstrate that the learned representations capture shared physical information. These modules also offer promising avenues for anomaly detection and similarity-based searches in large spectral archives.
\end{itemize}

Looking ahead, we plan to extend this framework to additional spectroscopic modalities, including LAMOST medium-resolution spectra (MRS, \citealp{2024ApJS..273...18L}), APOGEE infrared spectra \citep{2017AJ....154...94M}, Subaru PFS spectra \citep{2014PASJ...66R...1T}, and DESI DR1 spectra \citep{2025arXiv250314745D}. Our approach can be readily adapted to new instruments by pre-training modality-specific encoders and aligning them with contrastive objectives and decoder structures developed in this work. In future iterations, we will also explore efficient adaptation via neural network adapters, enabling scalable multi-survey alignment with minimal computational cost. In forthcoming work, a large-scale application and catalog release is planned.

 \section*{Acknowledgements} 
The Guoshoujing Telescope (the Large Sky Area Multi-Object Fiber Spectroscopic Telescope, LAMOST) is a National Major Scientific Project built by the Chinese Academy of Sciences. Funding for the project has been provided by the National Development and Reform Commission. LAMOST is operated and managed by the National Astronomical Observatories, Chinese Academy of Sciences.

Y.H. acknowledges support from the National Science Foundation of China (NSFC grant No. 12422303), the Fundamental Research Funds for the Central Universities (grant Nos. 118900M122, E5EQ3301X2, and E4EQ3301X2), and the National Key R\&D Program of China (grant No. 2023YFA1608303). X.Z. acknowledges partial support through a grant from the Schmidt Sciences Foundation. While this project was initiated prior to his current appointment at JHU, a significant portion of the work was completed during his Schmidt-supported position. 
T.C.B. acknowledges partial support from grants PHY 14-30152; Physics Frontier Center/JINA Center for the Evolution of the Elements (JINA-CEE), and OISE-1927130; The International Research Network for Nuclear Astrophysics (IReNA), awarded by the US National Science Foundation, and DE-SC002312; the Center for Nuclear Astrophysics Across Messengers (CeNAM), awarded by the U.S. Department of Energy, Office of Science, Office of Nuclear Physics. 
Y.S.T. is supported by the National Science Foundation under Grant No. AST-2406729. G.X. and X.T. acknowledge the support from Key R\&D Program of Zhejiang (2024SSYS0006).

We are thankful for useful discussions with Ce Sui, Alexander S. Szalay, Rosemary F.G. Wyse, and Benjamin D. Wandelt during the early stages of this work.

\section*{Data Availability}
All observational data used in this work are publicly available from the archives.  
A frozen version of the SpecCLIP software used in this analysis has been
archived on Zenodo \citep{zhao_2025_17824840} in compliance with the
AAS Journals software policy.  
No proprietary data were used.

\software{
    SpecCLIP \citep{zhao_2025_17824840},
    PyTorch \citep{NEURIPS2019_9015},
    Astropy \citep{2013A&A...558A..33A, 2018AJ....156..123A}
}
 
\vfill\eject
\appendix
\restartappendixnumbering

Here we present additional information on multiple aspects of this work, including elemental-abundance prediction, continuum fitting, normalization flows for parameter estimation, the use of pre-trained models, projection models and decoders, and loss curves. 

\section{Additional Results for Elemental-abundance Estimation}
\label{sec:pred_more}

\begin{figure}[htbp]
    \centering
    \begin{nolinenumbers}
    \begin{tikzpicture}
        \node[anchor=south west,inner sep=0] (main) at (0,0) {
            \includegraphics[width=0.95\textwidth]{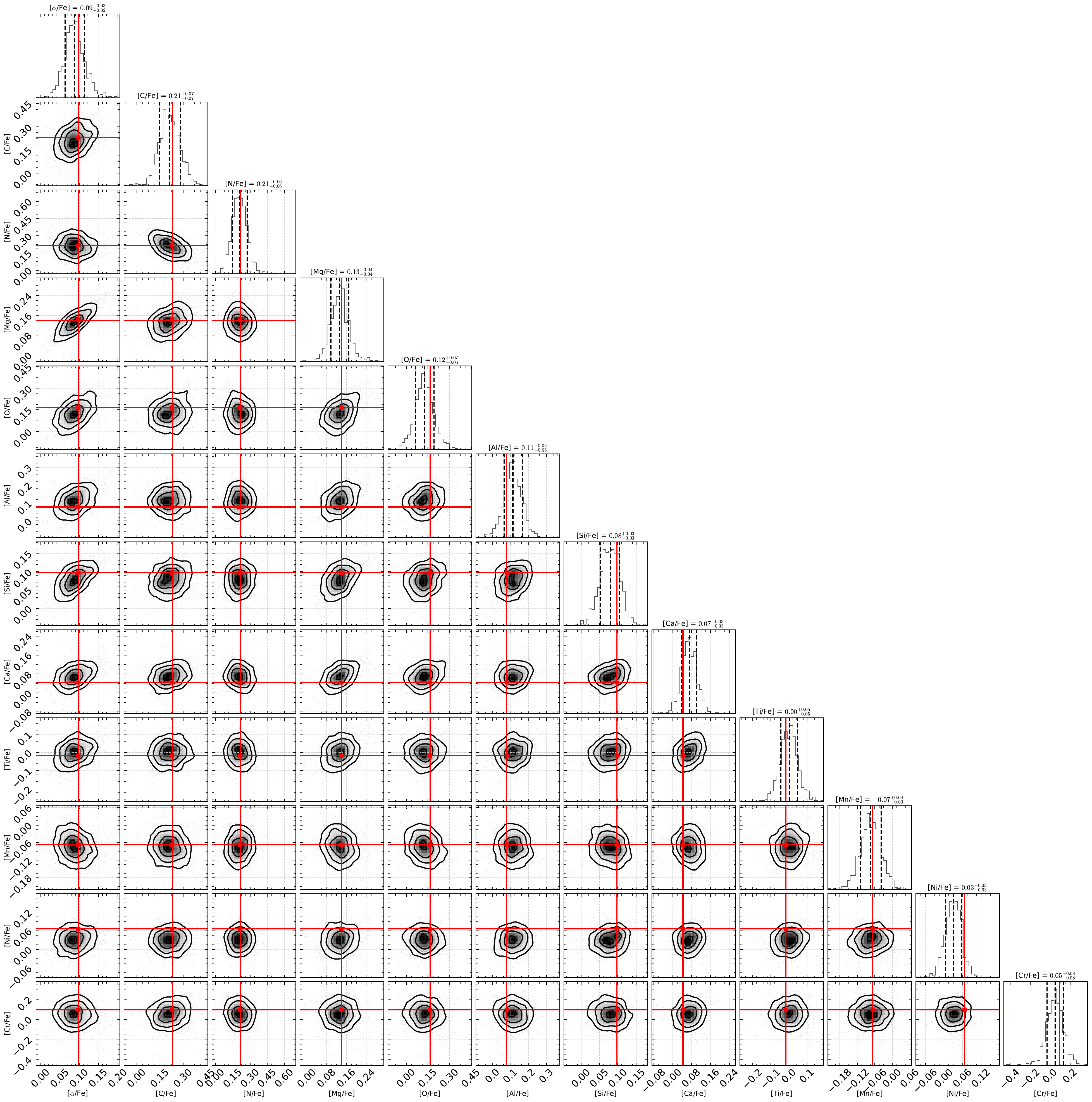}
        };
        
        \node[anchor=north east,inner sep=0] at (main.north east) {
            \includegraphics[width=0.5\textwidth]{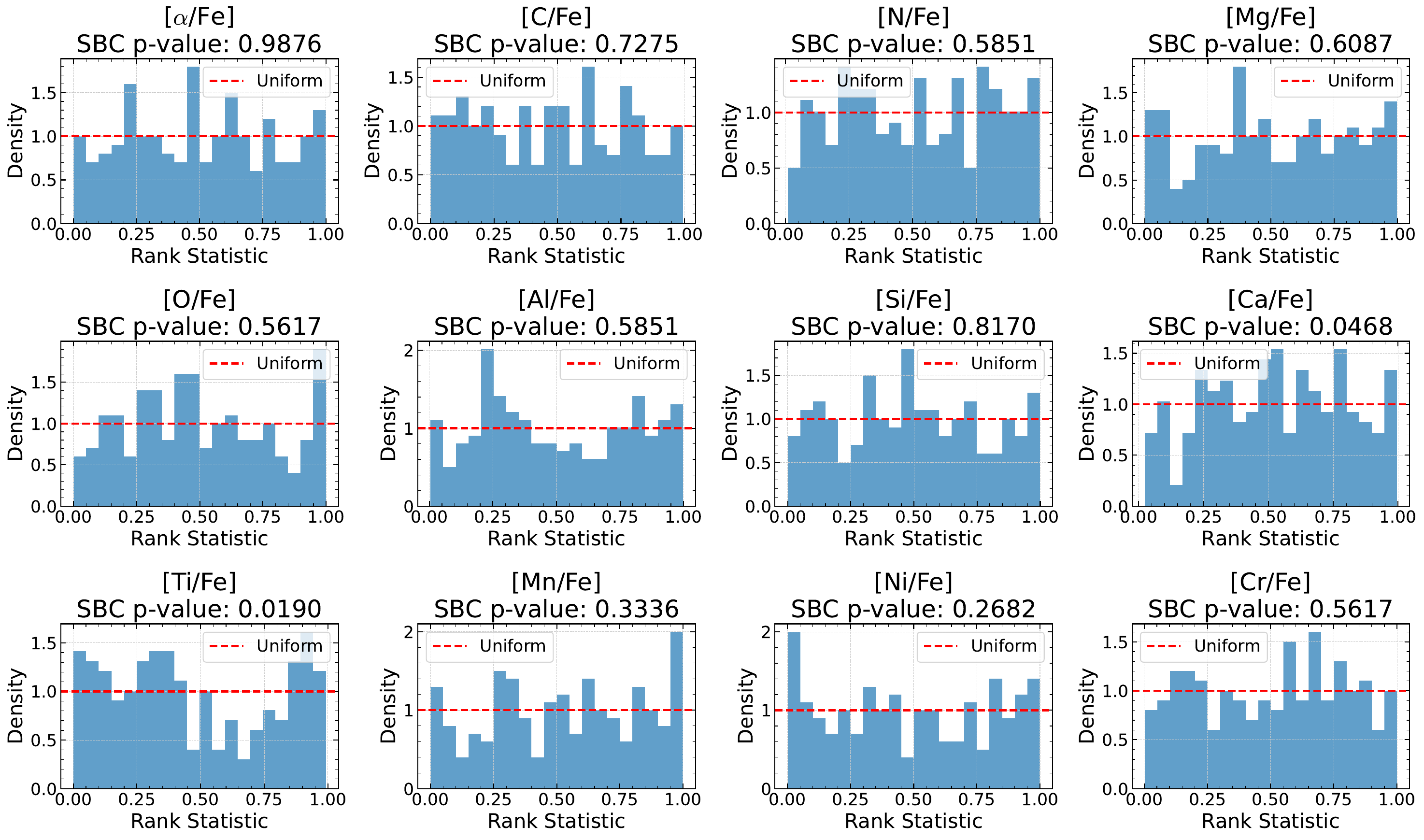}
        };
    \end{tikzpicture}
    \end{nolinenumbers}
    \caption{An example of posterior distributions of 12 elemental abundances inferred using simulation-based inference (SBI), with a downstream model trained jointly in the 12-dimensional parameter space. The embeddings used for training are from the pre-trained LRS foundation model. The upper-right panel shows simulation-based calibration (SBC) results, indicating that most posteriors are well-calibrated except for [Ti/Fe] and [Ca/Fe], which fall below the 0.05 significance threshold.}
    \label{fig:results_lrs_abundance}
\end{figure}

\begin{table*}[ht]
\centering
\scriptsize
\caption{Similar to Table~\ref{tab:compare_updated} for ``LRS Models'', but the results for all CLIP-based models are from the alignment between the LAMOST LRS MT and Gaia XP MT (see Section~\ref{sec:transformer_vs_oae} for discussion, and Appendices~\ref{sec:pre-trained} and~\ref{sec:proj_decode} for model details).}
\label{tab:compare_updated_lrs_mlp}
\begin{tabular}{lcccccc}
\hline
\multicolumn{7}{c}{\textcolor{red}{\textbf{LRS Models}}} \\
\hline
Parameter & Raw Spectra & Pre-trained & CLIP & CLIP-p & CLIP-pr & CLIP-split \\
 & $\sigma$ / $R^2$ & $\sigma$ / $R^2$ & $\sigma$ / $R^2$ & $\sigma$ / $R^2$ & $\sigma$ / $R^2$ & $\sigma$ / $R^2$ \\
\hline
\multicolumn{7}{l}{\textit{Atmospheric Parameters}} \\
$\mathrm{[Fe/H]}$ & 0.070 / $-$0.882 & 0.066 / 0.939 & 0.058 / 0.948 & 0.059 / 0.948 & \textbf{0.057} / 0.949 & 0.058 / \textbf{0.950} \\
$T_{\rm eff}$ (K)& 225.733 / 0.863 & 147.344 / 0.989 & 139.242 / 0.989 & \textbf{129.569} / \textbf{0.990} & 133.980 / \textbf{0.990} & 131.461 / \textbf{0.990} \\
$T_{\rm eff}$-sbi (maf) (K)& 106.903 / 0.979 & 94.942 / \textbf{0.990} & 97.396 / \textbf{0.990} & 95.147 / 0.989 & 96.208 / \textbf{0.990} & \textbf{93.155} / \textbf{0.990} \\
$T_{\rm eff}$-sbi (nsf) (K)& 76.986 / 0.982 & 84.991 / 0.991$^\dag$ & 86.515 / 0.991 & 86.133 / 0.989 & \textbf{84.517} / 0.990 & 84.694 / \textbf{0.992} \\
$\log g$ & 0.101 / 0.958 & 0.091 / 0.981 & 0.087 / 0.982 & 0.084 / 0.982 & 0.084 / 0.982 & \textbf{0.083} / \textbf{0.983} \\
$\log g$-sbi(maf) & 0.063 / 0.967 & \textbf{0.062} / 0.981 & 0.064 / \textbf{0.985} & 0.065 / 0.983 & 0.064 / 0.983 & 0.064 / 0.983 \\
\hline
\multicolumn{7}{l}{\textit{Elemental Abundances}} \\
$\mathrm{[\alpha/Fe]}$ & 0.023 / 0.872 & 0.021 / 0.906 & \textbf{0.020} / 0.912 & \textbf{0.020} / 0.912 & \textbf{0.020} / \textbf{0.913} & 0.021 / 0.909 \\
$\mathrm{[C/Fe]}$ & 0.041 / 0.758 & 0.039 / 0.792 & \textbf{0.038} / \textbf{0.806} & \textbf{0.038} / 0.804 & \textbf{0.038} / 0.805 & \textbf{0.038} / 0.803 \\
$\mathrm{[N/Fe]}$ & 0.054 / 0.598 & 0.052 / 0.642 & \textbf{0.050} / 0.662 & \textbf{0.050} / 0.661 & \textbf{0.050} / \textbf{0.665} & 0.051 / 0.657 \\
$\mathrm{[Al/Fe]}$ & 0.049 / 0.691 & 0.048 / 0.711 & \textbf{0.046} / 0.739 & \textbf{0.046} / 0.739 & \textbf{0.046} / \textbf{0.740} & \textbf{0.046} / 0.737 \\
$\mathrm{[Ca/Fe]}$ & 0.032 / 0.670 & 0.030 / 0.697 & \textbf{0.029} / 0.714 & \textbf{0.029} / 0.714 & \textbf{0.029} / \textbf{0.716} & 0.030 / 0.711 \\
$\mathrm{[Mg/Fe]}$ & \textbf{0.031} / 0.866 & 0.032 / 0.871 & \textbf{0.031} / \textbf{0.882} & \textbf{0.031} / \textbf{0.882} & \textbf{0.031} / 0.881 & \textbf{0.031} / 0.878 \\
$\mathrm{[Si/Fe]}$ & 0.029 / 0.776 & 0.029 / 0.803 & \textbf{0.028} / 0.814 & \textbf{0.028} / 0.813 & \textbf{0.028} / \textbf{0.816} & \textbf{0.028} / 0.807 \\
$\mathrm{[Ti/Fe]}$ & 0.061 / 0.492 & 0.058 / 0.532 & \textbf{0.056} / 0.551 & \textbf{0.056} / 0.552 & \textbf{0.056} / \textbf{0.555} & \textbf{0.056} / 0.544 \\
$\mathrm{[Mn/Fe]}$ & 0.033 / 0.761 & 0.032 / 0.780 & \textbf{0.031} / \textbf{0.800} & \textbf{0.031} / 0.799 & \textbf{0.031} / 0.798 & \textbf{0.031} / 0.792 \\
$\mathrm{[Ni/Fe]}$ & 0.027 / 0.426 & 0.026 / 0.454 & \textbf{0.025} / 0.487 & \textbf{0.025} / \textbf{0.489} & \textbf{0.025} / \textbf{0.489} & 0.026 / 0.479 \\
$\mathrm{[O/Fe]}$ & 0.051 / 0.698 & 0.050 / 0.722 & \textbf{0.049} / 0.728 & \textbf{0.049} / \textbf{0.729} & \textbf{0.049} / \textbf{0.729} & \textbf{0.049} / 0.728 \\
$\mathrm{[Cr/Fe]}$ & 0.081 / 0.177 & 0.076 / 0.225 & \textbf{0.075} / 0.234 & \textbf{0.075} / 0.233 & \textbf{0.075} / \textbf{0.237} & \textbf{0.075} / 0.232 \\
\hline
\multicolumn{7}{l}{\textit{Other Parameters}} \\
$E(BP-RP)$ & 0.075 / $-$36.886 & 0.076 / 0.711 & 0.070 / 0.746 & \textbf{0.069} / \textbf{0.748} & \textbf{0.069} / \textbf{0.748} & 0.070 / 0.742 \\
$v_{r}$ (km~s$^{-1}$) & 6.071 / 0.970 & \textbf{5.345} / 0.978 & 6.785 / 0.969 & 6.834 / 0.969 & 6.226 / 0.973 & 5.361 / \textbf{0.979} \\
$v_{r}$-sbi(maf) (km~s$^{-1}$)& \textbf{4.573} / 0.963 & 4.653 / 0.979 & 5.636 / 0.958 & 5.621 / 0.963 & 5.070 / 0.965 & 4.578 / \textbf{0.981} \\
\hline
\end{tabular}
\begin{flushleft}
\textit{Note.} 
Numbers in bold indicate the best performance (i.e., lowest $\sigma$ or highest $R^2$) for each parameter across all models. $^\dag$Results marked exclude a failed NSF sampling case on one extreme spectrum.
\end{flushleft}
\end{table*}

\begin{figure*}
\begin{center}
\includegraphics[scale=0.36,angle=0]{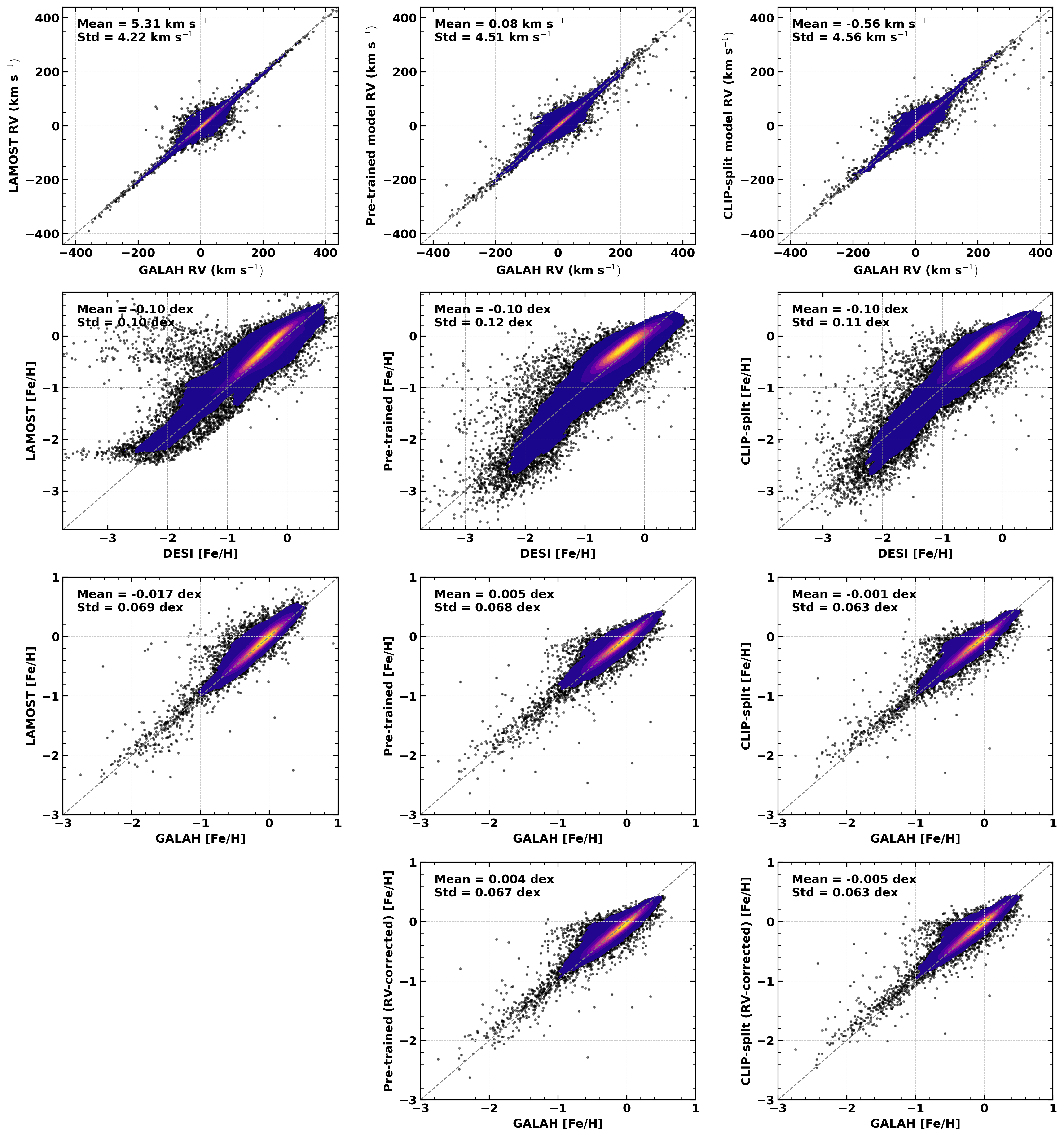}
\caption{Similar to Figure~\ref{fig:results_lrs_feh}, but the results for all CLIP-based models are from the alignment between the LAMOST LRS MT and Gaia XP MT (see Section~\ref{sec:transformer_vs_oae} for discussion, and Appendices~\ref{sec:pre-trained} and~\ref{sec:proj_decode} for model details).}
\label{fig:results_lrs_feh_mt}
\end{center}
\end{figure*}

Figure~\ref{fig:results_lrs_abundance} presents an example of chemical-abundance predictions from one input spectrum. This result is obtained using SBI trained with a single model for all parameters simultaneously. The inset in the upper right corner shows results from simulation-based calibration (SBC). Most elemental abundances exhibit well-calibrated posteriors, except for [Ti/Fe] and [Ca/Fe], which fall below the significance threshold of 0.05. This provides an example where, despite the model learns posterior distributions well overall, further tuning of the SBI architecture or its hyper-parameters is necessary to achieve reliable multivariate inference of elemental abundances--an aspect we did not further refine in this work.

In Table~\ref{tab:compare_updated_lrs_mlp}, we provide downstream task results using CLIP-based models aligned between the LAMOST LRS MT and Gaia XP MT encoders. This serves as a comparison to Table~\ref{tab:compare_updated}, where alignment is performed between the LAMOST LRS MT and Gaia XP OAE models. We find that the latter configuration shows overall comparable or better performance across the evaluation metrics, possibly due to the stronger pretraining of the XP OAE model. For completeness, in Figure~\ref{fig:results_lrs_feh_mt} we include an external comparison analogous to Figure~\ref{fig:results_lrs_feh} in the main text, but using the models from Table~\ref{tab:compare_updated_lrs_mlp}.

\section{Continuum Fitting Algorithm}
\label{sec: normalization}

Although we applied continuum fitting only to the blue segment ($4000\,\text{\AA} \leq \lambda \leq 5600\,\text{\AA}$) of the LAMOST LRS spectra for our analysis, we describe here the full continuum fitting algorithm, which is designed to robustly estimate the stellar continuum over the full wavelength range ($3850\,\text{\AA} \leq \lambda \leq 9000\,\text{\AA}$), and remains effective under varying SNRs. The method takes as input the observed wavelength array $\mathbf{w}$, the corresponding flux array $\mathbf{f}$, and an estimate of the average SNR, and returns a smooth continuum model $\mathbf{c}$.

The procedure is summarized as follows:

\begin{enumerate}
    \item \textbf{Pre-processing:}
    The flux array is first smoothed using a median filter of width 7 pixels to reduce the impact of narrow-line features and noise. The resulting smoothed flux is split into two wavelength segments: a blue side ($3700 \leq \lambda \leq 5700\,\text{\AA}$) and a red side ($\lambda > 6100\,\text{\AA}$).

    \item \textbf{Denoising with Savitzky–Golay filter:}
    The Savitzky–Golay filter is applied to each segment independently, with a smoothing window size of 3.

    \item \textbf{Blue Segment Adjustment (if sufficiently sampled):}
    A fifth-order polynomial is initially fit to the smoothed blue-side flux to identify the peak region. If the peak occurs at wavelengths $< 4500\,\text{\AA}$, interpolation is applied over three manually selected continuum windows ([4030--4160], [4270--4410], [4800--4940]\,\text{\AA}) to estimate local maxima and reduce the influence of absorption features on the continuum estimation. An iterative process is then performed. In each iteration, a fifth-order polynomial is fit to the updated flux, and the fit is used to suppress absorption features, effectively lifting the continuum. Ten such iterations are performed to ensure convergence.

    \item \textbf{Red Segment Correction:}
    A fourth-order polynomial fit is applied iteratively to the red-side segment. At each step, outliers deviating more than $3\sigma$ below the fit (or any point below the fit) are replaced by the polynomial value. This process is repeated up to 8 iterations to suppress absorption features and stabilize the continuum estimate.

    \item \textbf{Final Continuum Assembly:}
    The fitted continuum segments are combined to form the full continuum model $\mathbf{c}$ over the input wavelength grid. Values of $\mathbf{c} \leq 0$ are replaced by 1.0 to ensure a strictly positive continuum.
\end{enumerate}

\section{Normalizing Flows for Parameter Estimation}
\label{sec:nf}

Normalizing flows provides a flexible and tractable approach to modeling complex probability distributions by transforming a simple base distribution through a sequence of invertible and differentiable mappings. In SBI, we use normalizing flows to learn an approximate posterior \( q_\phi(\boldsymbol{\theta} \mid \mathbf{x}) \), conditioned on observations \( \mathbf{x} \), following the Neural Posterior Estimation (NPE) framework.

Let \( \mathbf{z} \sim p_Z(\mathbf{z}) \) denote a sample from a base distribution (e.g., standard Gaussian), and let \( f_\phi(\cdot; \mathbf{x}) \) be an invertible transformation conditioned on \( \mathbf{x} \), mapping \( \mathbf{z} \mapsto \boldsymbol{\theta} \). Then, the density of \( \boldsymbol{\theta} \) under the flow model is given by the change-of-variables formula:
\begin{equation}
    q_\phi(\boldsymbol{\theta} \mid \mathbf{x}) = p_Z(f^{-1}_\phi(\boldsymbol{\theta}; \mathbf{x})) \left| \det \left( \frac{\partial f^{-1}_\phi(\boldsymbol{\theta}; \mathbf{x})}{\partial \boldsymbol{\theta}} \right) \right| = p_Z(\mathbf{z}) \left| \det \left( \frac{\partial f_\phi(\mathbf{z}; \mathbf{x})}{\partial \mathbf{z}} \right) \right|^{-1},
\end{equation}
where \( \boldsymbol{\theta} = f_\phi(\mathbf{z}; \mathbf{x}) \).

For simplicity, we present the flow as a single transformation \( f_\phi \), but in practice it consists of a sequence of transformations:
\begin{equation}
    f_\phi = f_K \circ f_{K-1} \circ \cdots \circ f_1,
\end{equation}
where each \( f_k \) is an invertible and differentiable function with a tractable Jacobian. The log-determinant of the full transformation is the sum of the log-determinants of each layer:
\begin{equation}
    \log \left| \det \left( \frac{\partial f_\phi}{\partial \mathbf{z}} \right) \right| = \sum_{k=1}^K \log \left| \det \left( \frac{\partial f_k}{\partial \mathbf{h}_{k-1}} \right) \right|,
\end{equation}
where \( \mathbf{h}_k = f_k(\mathbf{h}_{k-1}) \), with \( \mathbf{h}_0 = \mathbf{z} \) and \( \mathbf{h}_K = \boldsymbol{\theta} \).

\subsection{Masked Autoregressive Flow (MAF)}
MAF \citep{2017arXiv170507057P}) models the forward transformation \( \mathbf{z} \mapsto \boldsymbol{\theta} \) as a sequence of autoregressive operations. Each component of \( \boldsymbol{\theta} \) is computed as:
\begin{equation}
    \theta_i = \mu_i(\mathbf{z}_{<i}; \mathbf{x}) + \sigma_i(\mathbf{z}_{<i}; \mathbf{x}) \cdot z_i,
\end{equation}
where \( \mu_i \) and \( \sigma_i \) are outputs of neural networks that depend on the previous components \( \mathbf{z}_{<i} \) (or equivalently, the previous output components $\boldsymbol{\theta}_{<i}$) and the conditioning variable \( \mathbf{x} \). The Jacobian of this transformation is lower triangular, which allows the log-determinant to be computed efficiently:
\begin{equation}
    \log \left| \det \left( \frac{\partial \boldsymbol{\theta}}{\partial \mathbf{z}} \right) \right| = \sum_i \log \sigma_i(\mathbf{z}_{<i}; \mathbf{x}).
\end{equation}

\subsection{Neural Spline Flow (NSF)}
NSF \citep{2019arXiv190604032D} generalizes MAF by replacing affine transformations with monotonic rational-quadratic splines. Each component is transformed as:
\begin{equation}
    \theta_i = \text{spline}_i(z_i; \psi_i(\mathbf{z}_{<i}, \mathbf{x})),
\end{equation}
where \( \psi_i \) are spline parameters (bin widths, heights, and derivatives) predicted by a neural network conditioned on \( \mathbf{z}_{<i} \) and \( \mathbf{x} \). The log-determinant of the Jacobian is given by:
\begin{equation}
    \log \left| \det \left( \frac{\partial \boldsymbol{\theta}}{\partial \mathbf{z}} \right) \right| = \sum_i \log \left| \frac{\partial \theta_i}{\partial z_i} \right|,
\end{equation}
where each derivative term is efficiently computed from the analytical form of the spline. Note that while this autoregressive form is analogous to MAF, the default implementation based on the provided code is often the coupling layer variant of NSF (NSF-C), which is typically more efficient for density evaluation. Details on the coupling layer structure can be referred to in \citet{2019arXiv190604032D}.

\medskip

In summary, both MAF and NSF enable flexible posterior approximation in the NPE setting, while maintaining exact likelihood evaluation and efficient training via maximum likelihood.

\section{Pre-trained Models}
\label{sec:pre-trained}

In this paper, we tried two different kinds of pre-trained models, one is the transformer-based model and the other is a MLP-based auto-encoder. Both kinds of networks have the same number of trainable parameters (42.7 million). We use a batch size of 64 per GPU. The models are optimized using AdamW \citep{2017arXiv171105101L}
with a learning rate of $1 \times 10^{-5}$ and a weight decay of 0.1. The learning rate follows a cosine annealing schedule with linear warm-up.

\subsection{Transformer-Based Spectral Pre-trained Models}

We describe two transformer models designed for the masked reconstruction of stellar spectra, based on \citet{2024MNRAS.531.4990P}, each tailored to the characteristics of a different input data set: LAMOST and Gaia XP.

\subsubsection{Masked Spectral Modeling for LAMOST Spectra}

This model is designed for higher-resolution (compared with Gaia XP) spectra from instruments like LAMOST, where each input sample is $x \in \mathbb{R}^{T \times 1}$ with $T=1462$ wavelength bins. The model operates as a masked sequence autoencoder using transformers.

\paragraph{Input Pre-processing.} 
Each spectrum $x \in \mathbb{R}^{T \times 1}$ is standardized with:
\begin{equation}
\quad \mu = \frac{1}{T} \sum_{t=1}^T x_t,\quad \sigma^2 = \frac{1}{T} \sum_{t=1}^T (x_t - \mu)^2
\end{equation}
Then, the standardized spectrum is sliced into overlapping chunks of length $L$ (e.g., $L=20$) with overlap $O=10$, forming an input sequence of length $S=146$, and a special token $x'_0 = \log_{10} \sigma$ is pre-pended, forming an extended sequence $x' \in \mathbb{R}^{T' \times (L+1)}$ with sequence length $T'=S+1$.

\paragraph{Model Architecture.}
The input is linearly embedded and added to the learned positional embeddings.
Then it passes through $N=6$ layers of standard transformer blocks with $H=6$ attention heads. The output is normalized and decoded through a linear projection.

\paragraph{Masking Strategy.}
To train the model in a self-supervised fashion, we applied a chunk-based masking strategy. Given an input sequence of length $T'$, we conceptually divide it into $M=6$ segments of equal-length and randomly select a contiguous piece of width $w=10$ within each segment to mask. For each chunk, its starting index is sampled uniformly from the allowable range within the segment. This ensures that the masked regions are distributed across the sequence and sufficiently separated.

Formally, let the $i$-th segment span the indices $[s_i, s_i + \ell]$, where $\ell = \lfloor T' / M \rfloor$. Then the masked region for segment $i$ is:
\begin{equation}
\text{Mask}_i = [s_i + \delta_i,\, s_i + \delta_i + w), \quad \delta_i \sim \mathcal{U}(0, \ell - w)
\end{equation}
The masked input $\tilde{x}$ is defined as:
\begin{equation}
\tilde{x}_t = 
\begin{cases}
0, & \text{if } t \in \cup_i \text{Mask}_i \\
x'_t, & \text{otherwise}
,
\end{cases}
\end{equation}
where $t \in [0,\ T'-1]$ is the sequence index. This strategy preserves long-range contextual integrity and forces the model to interpolate realistic spectral values across variable scales. Compared to random masking, chunk-based masking is more appropriate for spectral data, where features span contiguous wavelength regions.

\paragraph{Loss Function.}
Let $f_\theta(\tilde{x})$ denote the model’s reconstruction output, the model is trained to reconstruct only the masked regions using a masked MSE loss:
\begin{equation}
\mathcal{L}_{\text{masked}} = \frac{1}{\sum_t m_t} \sum_{t=1}^{T'} m_t \left\| f_\theta(\tilde{x})_t - x'_t \right\|^2
\end{equation}
where $m_t = 1$ if $t$ is masked and $0$ otherwise.

LAMOST spectra benefit from local patterns and detailed features. Thus, using dense chunk-based masking and slicing helps the transformer leverage locality while preserving the global context. We train this model with 8 GPUs 
over a total of 128 epochs, a process that takes roughly 20 hours.

\subsubsection{Masked Spectral Modeling for Gaia XP Spectra}

This model targets low-resolution Gaia XP spectra where each sample is $x \in \mathbb{R}^{T \times 1}$ with $T=343$ wavelength bins.

\paragraph{Input Pre-processing.}
As in the LAMOST model, we standardize the spectrum and pre-pend two  (mean and standard deviation) tokens:
\begin{equation}
x'_0 = \mu,\quad x'_1 = \sigma
\end{equation}
forming an extended sequence $x' \in \mathbb{R}^{(T+2) \times 1}$.

The model architecture and the loss function are similar to the LAMOST case.

We train this model with 8 GPUs and a total of 191 epochs, which is roughly 40 hours.

\subsection{MLP-based Autoencoder}
\label{sec:mlp_based_ae}

For comparison with the transformer-based models, we also construct an MLP-based autoencoder to pre-train the stellar spectra. For both LAMOST LRS and Gaia XP, the network architectures are similar: an initial projection layer of shape $[\text{input\_dim}, \text{hidden\_dim}]$; the encoder consists of two MLP blocks, each with structure $[\text{hidden\_dim}, 3 \times \text{hidden\_dim}, \text{hidden\_dim}]$; followed by a bottleneck layer of shape $[\text{hidden\_dim}, 768]$. The decoder mirrors this structure to reconstruct the input spectra. For LAMOST LRS, we use $\text{input\_dim} = 1462 + 1$ and $\text{hidden\_dim} = 1245$; for Gaia XP, we use $\text{input\_dim} = 343 + 2$ and $\text{hidden\_dim} = 1290$. The former includes one additional (logarithmic) standard deviation as the first token, while the latter includes both the mean and the standard deviation as the first two tokens. We train the LRS model on 4 GPUs 
for 128 epochs, taking approximately 140 minutes, and the Gaia XP model 4 GPUs for 191 epochs, taking approximately 150 minutes.

\section{Projection Models and Decoders}
\label{sec:proj_decode}

This section provides details about the modules used in CLIP-based model training. For these models, we use 8 GPU and require a total of roughly 3 hours for training. All model variants use a batch size of 1024 per GPU for contrastive training with decoders. We adopt the AdamW optimizer with a learning rate of $1 \times 10^{-4}$ and a weight decay of 0.05. The learning rate is scheduled using a cosine annealing schedule with linear warm-up.

\subsection{Projection Networks}

For the projection networks, the LAMOST LRS model follows \citet{2024MNRAS.531.4990P}. We first obtain the output from a pre-trained LAMOST LRS model, then go through a cross-attention module with a learnable query vector. The cross-attention module has four attention heads with an output dimension of 768. Finally, we have an MLP layer with hidden features that have dimension $4 \times 768 $. We did not compress the information more in this projected network in order to investigate the information gain without compression. Therefore, the dimension of the final projected embedding is still 768. The number of trainable parameters is about 7.1 million.

For the Gaia XP spectra pre-trained with the transformer-based spectral model, we use the same projection networks as for the LAMOST LRS model. For the Gaia XP spectra pre-trained with a MLP-based autoencoder, our projection network has the dimension of $[768, 768, 1160, 768]$, with a residual MLP block at the end with hidden dimension $4 \times 768$. The choice of number of layers and the dimension of each layer is arbitrary; the key control condition is to maintain the same number of total trainable parameters as its cross-attention counterpart (7.1 million). 

For the CLIP-split model, the LAMOST LRS projection network (5.1 million) outputs two branches for shared and non-shared representations:
    \begin{itemize}
        \item \textbf{Shared}: uses \texttt{CrossAttentionHead} with 512-D projection and $n=4$ heads.
        \item \textbf{Non-shared}: uses \texttt{CrossAttentionHead} with 256-D projection and $n=2$ heads.
    \end{itemize}
Both branches use MLPs with hidden size $4 \times 768$.

The Gaia XP projection network paired with the transformer-based pre-trained model has the same architecture as the LAMOST LRS projection network.  For the projection network (also 5.1 million trainable parameters) paired with MLP-based pre-trained model, it outputs two latent representations:
    \begin{itemize}
        \item \textbf{Shared}: linear projection to 512-D, followed by MLP: [512, 1160, 512].
        \item \textbf{Non-shared}: linear projection to 256-D, followed by MLP: [256, 1160, 256].
    \end{itemize}
Each pathway includes a residual MLP block at the end with hidden dimension of $4 \times 512$ and $4 \times 256$, respectively.

\subsection{Decoders}

For the CLIP-pr model, the XP Decoder and LRS $\rightarrow$ XP cross decoder share the same architecture with MLP layer dimensions: [in, 4$\times$in, 2$\times$in, in, out] (8.5 million). The LRS decoder and XP $\rightarrow$ LRS cross decoder share the same architecture with layer dimensions: [in, 4$\times$in, 4$\times$in, 4$\times$in, out] (25.8 million), where in $=768$; out $=1462$ and $343$ for LAMOST LRS and Gaia XP, respectively. Note that for LAMOST LRS, we are only reconstructing the normalized spectra with the mean and standard deviation calculated for each spectra, similar for the CLIP-split model.

For the CLIP-split model, both LRS (14.0 million) and XP (0.1 million) decoders take shared and non-shared features to reconstruct the original spectra. Shared and non-shared inputs are projected to the out dimension, where out $=1462$ and $343$ for LAMOST LRS and Gaia XP, respectively. The two outputs are concatenated, and the final reconstruction layer contains [out$\times$2, out$\times$2, out].

For the cross-modal decoder of the CLIP-split model, we take only the shared representation as input. The decoder architecture depends on the output dimension, yielding approximately 12.5 and 3.9 million trainable parameters for the following two setups:
    \begin{itemize}
        \item \textbf{For LAMOST (out $>$ shared)}: [512, 2048, 2048, 2048, out].
        \item \textbf{For Gaia XP (out $\leq$ shared)}: [512, 2048, 1024, 512, out].
    \end{itemize}

\section{Loss Curves}
\label{sec:loss_curve}

\begin{figure*}
\begin{center}
\includegraphics[scale=0.35,angle=0]
{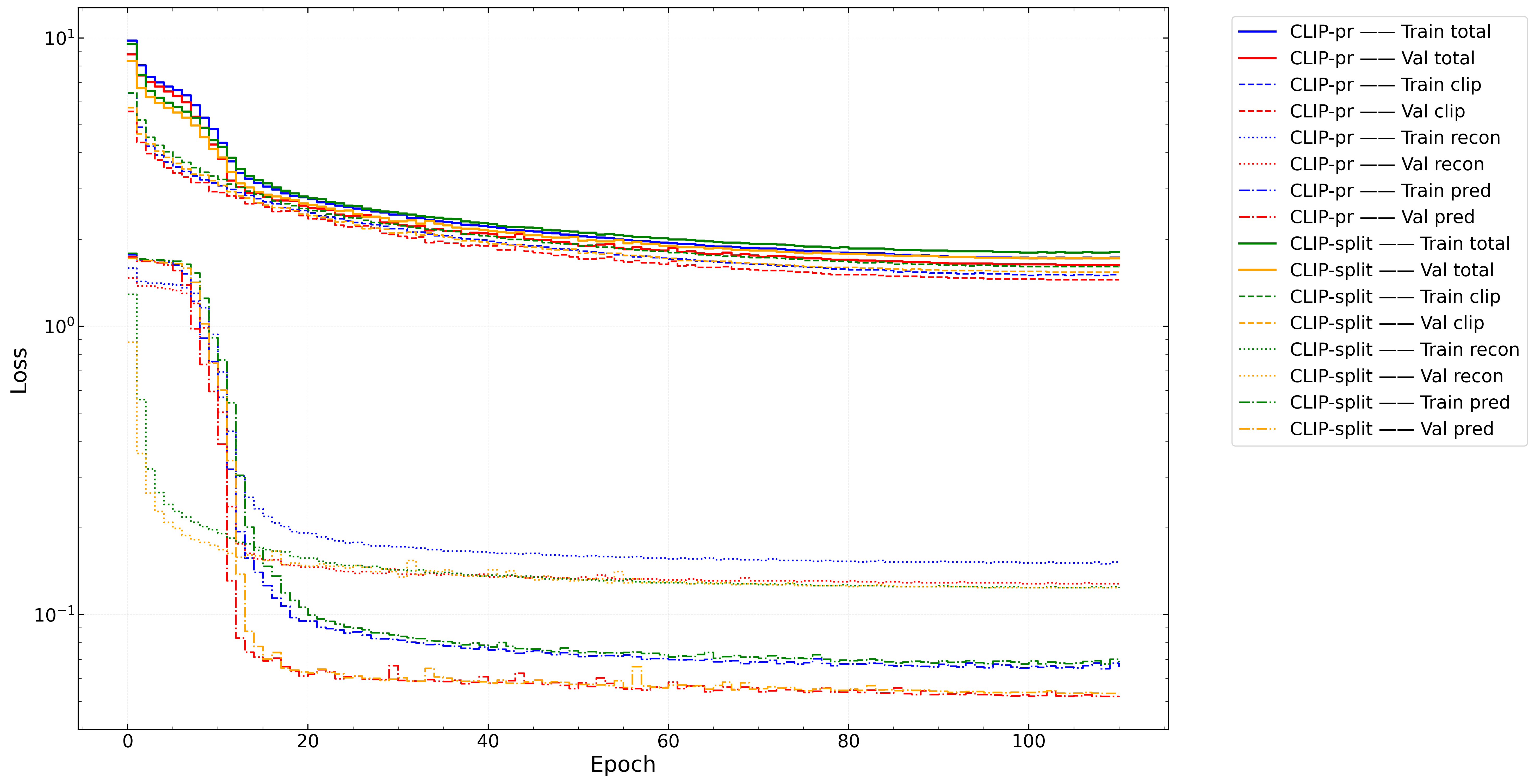}
\caption{Loss curves of the CLIP-pr and CLIP-split models.}
\label{fig:loss}
\end{center}
\end{figure*}

This section presents the comparisons of the loss curves between the CLIP-pr and CLIP-split models, as shown in Figure~\ref{fig:loss}. Although the CLIP-pr model achieves a lower CLIP loss during training, it exhibits a lower absolute cosine similarity in test pairs, as shown in Table~\ref{tab:compare_similarity_pred_score}. This discrepancy may arise from two factors. First, the geometry of the high-dimensional embedding spaces—768 dimensions in the CLIP-pr projected embedding versus 512 in the CLIP-split shared embedding—tends to yield lower cosine similarity in the former. Second, the CLIP loss focuses on relative alignment rather than absolute similarity. Additionally, the CLIP-pr model yields slightly lower cross-modal prediction losses. In contrast, the CLIP-split model converges more quickly in learning the reconstruction and achieves marginally lower reconstruction loss, likely due to its design of a non-shared projected embedding space.

\section{Summary of Key Hyper-parameters and Configurations}
\label{sec:hyper_summary}

Table~\ref{tab:hyperparams_combined} summarizes the key hyper-parameters and configurations used across the main training stages. 
For complete details and implementation settings, please refer to Appendix~\ref{sec:pre-trained}, Appendix~\ref{sec:proj_decode}, and Section~\ref{sec:clip_intro}.

\begin{sidewaystable*}[ht!]
\centering
\caption{Summary of hype-parameters and configurations across main training stages.}
\label{tab:hyperparams_combined}
\renewcommand{\arraystretch}{1.1}
\footnotesize

\begin{minipage}{0.9\textheight}
\centering
\begin{tabular}{lccccc}
\hline
\textbf{Setting} & \textbf{LRS MT} & \textbf{XP OAE} & \textbf{CLIP} & \textbf{CLIP-r} & \textbf{CLIP-p} \\
\hline
Encoder / Layers & 6 Self-Attn (6 heads) & 2 MLP blocks enc.+dec. & Cross-Attn (4 heads)+MLP & Cross-Attn (4 heads)+MLP & Cross-Attn (4 heads)+MLP \\
Latent dim. & 768 & 768 & 768 & 768 & 768 \\
Tokenization & Chunks ($L{=}20$, $O{=}10$)+$\log_{10}\sigma$ & Prepend $\mu,\sigma$ & Use pre-trained embeddings & Same as CLIP & Same as CLIP \\
Masking / Rate & Chunk-based ($\sim$45\%) & – & – & – & – \\
Optimizer & AdamW & AdamW & AdamW & AdamW & AdamW \\
Learning rate & $1\times10^{-5}$ & $1\times10^{-5}$ & $1\times10^{-4}$ & $1\times10^{-4}$ & $1\times10^{-4}$ \\
Weight decay & 0.10 & 0.10 & 0.05 & 0.05 & 0.05 \\
LR schedule & Cosine + warm-up & Cosine + warm-up & Cosine + warm-up & Cosine + warm-up & Cosine + warm-up \\
Batch / GPU & 64 & 64 & 1024 & 1024 & 1024 \\
GPUs & 8 & 4 & 8 & 8 & 8 \\
Epochs / Time & 128 / 20 h & 191 / 2.5 h & 110 /3 h & 110 / 3 h & 110 / 3 h \\
Loss fn. & Masked MSE & MSE & $\mathcal{L}_{\text{CLIP}}$ & $\mathcal{L}_{\text{CLIP}}+w_\text{recon}\mathcal{L}_{\text{recon}}$ & $\mathcal{L}_{\text{CLIP}}+w_\text{pred}\mathcal{L}_{\text{pred}}$ \\
Loss weights & – & – & – & $w_\text{recon}=1.0$ & $w_\text{pred}=1.0$ \\
CLIP temp $\tau$ & – & – & 15.5 & 15.5 & 15.5 \\
\hline
\end{tabular}
\end{minipage}

\vspace{1.5em}

\begin{minipage}{0.9\textheight}
\centering
\textbf{(Continue)}\\[4pt]
\begin{tabular}{lccccc}
\hline
\textbf{Setting} & \textbf{CLIP-pr} & \textbf{CLIP-split} & \textbf{Downstream MLP} & \textbf{Downstream SBI (NPE)} \\
\hline
Encoder / Layers & Cross-Attn (4 heads)+MLP & 2 Cross-Attn branches & 4 MLP [in,1024,512,64,1] & 2 transforms, 60 hidden units each \\
Latent dim. & 768 & 512+256 & – & – \\
Tokenization & Same as CLIP & Same as CLIP & Raw spectra / embeddings & Raw spectra / embeddings \\
Masking / Rate & – & – & – & – \\
Optimizer & AdamW & AdamW & AdamW & Adam  \\
Learning rate & $1\times10^{-4}$ & $1\times10^{-4}$ & $1\times10^{-5}$ & $5\times10^{-4}$ \\
Weight decay & 0.05 & 0.05 & $1\times10^{-4}$ & – \\
LR schedule & Cosine + warm-up & Cosine + warm-up & ReduceLROnPlateau & – \\
Batch / GPU & 1024 & 1024 & 32 & 50 \\
GPUs & 8 & 8 & 1 & 1 \\
Epochs / Time & 110 / 3 h & 110 / 3 h & Max. 100 / $<$10 min & – / $<$10 min \\
Loss fn. & $\mathcal{L}_{\text{CLIP}}+w_\text{recon}\mathcal{L}_{\text{recon}}+w_\text{pred}\mathcal{L}_{\text{pred}}$ & $\mathcal{L}_{\text{CLIP}}+w_\text{recon}\mathcal{L}_{\text{recon}}+w_\text{pred}\mathcal{L}_{\text{pred}}$ & MSE & Negative log-likelihood \\
Loss weights & $w_\text{recon}=w_\text{pred}=1.0$ & $w_\text{recon}=w_\text{pred}=1.0$ & – & – \\
CLIP temp $\tau$ & 15.5 & 15.5 & – & – \\
\hline
\end{tabular}
\end{minipage}

\end{sidewaystable*}

\vfill\eject
\bibliography{sppara_calib}{}
\bibliographystyle{aasjournal}
\end{document}